\documentclass[11pt]{article}

\usepackage[utf8]{inputenc}
\usepackage[T1]{fontenc}
\usepackage{lmodern}
\usepackage[a4paper,margin=1in]{geometry}
\usepackage{microtype}
\usepackage[numbers,sort&compress]{natbib}

\usepackage{amsmath,amssymb,amsthm,mathtools}
\usepackage{graphicx}
\usepackage{hyperref}
\usepackage[capitalize]{cleveref}

\newcommand{\namedcat}[1]{\mathsf{#1}}
\newcommand{\C}{\namedcat{C}}
\newcommand{\X}{\namedcat{X}}
\newcommand{\A}{\namedcat{A}}
\newcommand{\Set}{\namedcat{Set}}
\newcommand{\Fin}{\namedcat{Fin}}

\newtheorem{theorem}{Theorem}

\theoremstyle{definition}
\newtheorem{definition}[theorem]{Definition}
\theoremstyle{remark}



\title{Compositional System Dynamics: The Higher Mathematics Underlying System Dynamics Diagrams \& Practice}

\author{Xiaoyan Li\thanks{Corresponding author: \href{mailto:xiaoyan.li2@uleth.ca}{xiaoyan.li2@uleth.ca}.}
\and Evan Patterson
\and Patricia L. Mabry
\and Nathaniel D. Osgood }

\date{} 

\begin{document}
\maketitle

\begin{abstract}
This work establishes a robust mathematical foundation for compositional System Dynamics modeling, leveraging category theory to formalize and enhance the representation, analysis, and composition of system models. Here, System Dynamics diagrams, such as stock \& flow diagrams, system structure diagrams, and causal loop diagrams, are formulated as categorical constructs, enabling scalable, transparent, and systematic reasoning. By encoding these diagrams as data using attributed C-sets and utilizing advanced categorical tools like structured cospans, pushouts, pullbacks, and functor mappings, the framework supports modular composition, stratification, and seamless mapping between syntax and semantics.

The approach underwrites traditional practice with firm mathematical structure, facilitates the identification of certain forms of pathways and feedback loops, the detection of simple patterns within complex diagrams, common structure between diagrams, and structure-preserving mappings between diverse diagram types. Additionally, this framework supports alternative semantics, such as stochastic transition dynamics, extending beyond traditional ordinary differential equation (ODE) representations. Applications in compositional modeling, modularity, and team-based collaboration demonstrate the practical advantages of this advanced framework.

Future directions include integrating dimensional annotations, supporting hybrid and agent-based modeling paradigms, and expanding the framework's applicability to global and local temporal reasoning through temporal sheaves. By revealing and formalizing the hidden mathematical structure of System Dynamics diagrams, this work empowers practitioners to tackle complex systems with clarity, scalability, and rigor.

\end{abstract}


% ...


\section{Introduction}\label{Introduction}

Compared to other simulation traditions, System Dynamics is distinguished by its diverse repertoire of diagrams, their accessibility to project stakeholders, and their central role in which they facilitate insights into system behavior.  A great deal of the value delivered by System Dynamics modeling project is realized by the accessibility and varied abstraction levels associated with such diagrams, and the different dynamic patterns that they help explain.

While creation, refinement of and thinking around such diagrams---whether stock \& flow models, system structure diagrams, and causal loop diagrams---constitutes the central activity of many modeling projects, there has been surprisingly little examination of deeper mathematical underlying System Dynamics diagrams, the formal relationships of different diagrams to each other, and formalization of the relationships between diagram types.  This work highlights the clear mathematical underpinnings for each such diagram type and which relate those types to each other.  In the process of doing so, it illustrates the compelling benefits for thinking and tool use to be secured by a truly compositional form of System Dynamics, particularly as it relates to scalability, transparency, performance, clarity, cohesiveness, productivity, and versatility of System Dynamics modeling. While deferring more detailed treatment to our recently published manuscripts on the mathematical structure of System Dynamics within the mathematics literature, this paper introduces key mathematical concepts that enable these advancements, illustrating their practical affordances and advantages. Additionally, we discuss ongoing and future work to extend this foundational approach to broaden its applicability and impact. Ultimately, this effort seeks to formalize and expand System Dynamics practice, offering a cohesive framework for models across different levels of abstraction and enhancing our ability to relate multiple diagrams and diagram types to each other, while maintaining fidelity to the respective roles of each diagram type within in system representation and System Dynamics practice.

\subsection{Motivation}

The motivation for this work stems from the prospect of facilitating systems thinking, System Dynamics model conceptualization and formulation, and model evolution and system understanding, by making the surrounding tools more powerful, integrated, and flexible.  The methods described here demonstrably realize such desiderata, but their implications are much broader-reaching.  Instead of aspiring to exhaustively enumerate the varied benefits, we elect here to characterize major motivations for this work from two angles:  1) At a concrete level, by envisioning some compelling use cases to be realized by these techniques 2) At a slightly more abstract level, by characterizing gaps and limitations with current approaches.  We first begin with some concrete use cases.

\subsubsection{Example Use Cases}

We describe here some ways in which the techniques here can supplement the repertoire of a System Dynamics modeler and team member.  We foresee practice where:
\begin{itemize}
    \item System Dynamics modelers can readily build up stock and flow diagrams or causal loop diagrams by gluing together modular, previously vetted or peer reviewed models.  For stock and flow diagrams, these pieces could further be pre-calibrated components, checked for unit consistency, tested with extreme values, and documented.
    \item Such models can be \textbf{composed} together in a variety of ways.  For stock and flow and system structure diagrams, this could involve joining models together by identifying variables within the associated diagrams -- treating such variables as the same in the composite model.  But it could also involve setting one diagram upstream of another, with a flow out of the upstream model being connected to the downstream model.  An alternative form of substitution would substitute one entire submodel to replace a single stock in another model, given certain compatibility criteria.  When both diagrams  are associated with dimensional/unit information, such composition techniques can be further facilitated by requiring variables being identified between the two models to have identical dimensions/units.
    \item A stock and flow diagram can be interpreted not only through the lens of simulation as underlying ordinary differential equations, but many other ``semantics'' as well -- types of simulation, but also structural analysis, dimensional scaling, and other purposes.  For example, that same stock and flow model could be interpreted such that designated stocks and flows are interpreted as stochastic transitions of discrete entities (e.g., people, doses of vaccine, support dogs, etc.), as characterized by a Poisson process, with a rate given by the formula for the flow. Alternatively, the stock and flow model could be interpreted to identify loop gain over time \cite{kampmann2012feedback}, or through the lens of eigenvalue elasticity analysis \cite{saleh2005comprehensive,guneralp2005progress,ZhangQianGlobalEigenvalueElasaticityAnalysis}.  Similarly, a causal loop diagram analysis could be interpreted through the lens of identifying betweenness centrality of variables, variable presence on feedback loops, etc.
    \item A modeler working with a stock and flow diagram can readily search for molecules of structure \cite{molecules, hines1996molecules} within that diagram, finding patterns such as first-order and higher-order delays, coflows, aging chains, positive and negative feedback loops; as above, dimensional information could be used to sharpen such motif-finding.  Similarly, a causal loop diagram modeler could search for instances of patterns, such as negative and positive feedback, and negative feedbacks with delays.
    \item Modelers working with a causal loop diagram can flexibly visualize various coarse-grainings of that diagram, to highlight or explore larger structural patterns.
    \item That modeler could further compare that causal loop diagram to another -- perhaps a variant or later version of the first -- to have a systematic characterization of how one relates to another.  The identified relationships would find cases where the first diagram was embedded in the other, coarse-graining of sections of the diagram in one diagram compared to the other, elaborations of causal pathways, where multiple variables map to a single variable, etc.
    \item A modeler can readily extract a causal loop diagram from a stock and flow diagram---deriving requisite polarities for connections---and compare that to a larger causal loop diagram constructed earlier.
    \item When building stock and flow diagrams seeking to incorporate larger amounts of heterogeneity, we can fairly modularly layer in individual layers of heterogeneity, in a manner that greatly lessens and localizes the work required, and allows for handling cases in which the dimensions of stratification are not fully independent. 
\end{itemize}

\subsubsection{Gaps and Limitations}

This work is motivated from the desire to enhance the ease and insightfulness of System Dynamics practice by using an integrated set of higher mathematical methods that make possible certain new types of analyses, simplify application of existing innovative methods in System Dynamics, and ease other common tasks.

While a great deal of benefit has been secured from effort devoted to understanding specific semantic domains---such as ordinary differential equations (ODEs)---by which we can interpret the dynamics of System Dynamics models \cite{sterman2000business}, formal characterization of the mathematical structure of the diagrams themselves has been comparatively neglected. This oversight hinders the rigorous exploration of key relationships within and across different types of diagrams, leaving unrealized certain opportunities for insight from the diagrams that make System Dynamics so powerful, and has complicated the task of conducting other forms of analysis.  We mention here a few examples of where a lack of a rigorous mathematical basis for System Dynamics diagrams shortchanges current practice.  While we recognize that some of the challenges below have been the target more informal methods---such as those rooted in \textit{ad hoc} manipulation in software---our goal here is to emphasize the breadth of benefits that a systematic, unifying, mathematical approach to formalizing can offer in addressing needs in System Dynamics practice.

One major resulting limitation involves barriers in formally relating instances of the same diagram type. For example, while informal recognition of points of similarity and difference between two diagrams of a given type is a common need within System Dynamics practice, there is no systematic and formal approach to systematically relate two stock \& flow diagrams at different levels of granularity or elaboration, or between two causal loop diagrams where one embeds, coarse-grains or further explicates another. 

Similarly, while it is routine for projects employing System Dynamics to maintain stock \& flow models and causal loop diagrams representing the same system at different levels of abstraction, and while informal comparison of one to the other---such as identifying common features, or areas reflecting different scope---are basic to contemporary practice, no established, mathematically rigorous framework to connect diagrams of distinct type. As with diagrams of the same type, this lack of a systematic mechanism for relating diagrams of different types is unwarranted, given the underlying mathematical relationships that exist between them. This lack of formalism isolates diagrams, leading them to be treated formally as solitary entities and undermining the potential for tools and analysis that exploit the ability to represent the system in a cohesive, multi-level way.

Another key challenge lies in the absence of a mathematical framework for model composition. In practical modeling scenarios, it is often necessary to combine different model components -- such as integrating a vaccination process model with a disease spread model, or linking representations of the aquatic and terrestrial life phases of mosquitoes in an ecological system. However, existing approaches offer little guidance on how to formally compose models, whether by connecting their variables, sharing stocks, or embedding one model's details within another.  While such connections can be manually achieved in software, the lack of a formal mathematical description imposes a number of costs.  For example, the absence of a formal framework leaves unclear how the structure of the composite model relates to that of the subpieces, which limits the ability to reason about how both the structural and dynamic properties of the pieces relate to those for the composite structure.  The absence of a formal compositional framework also shortchanges the ability for software to facilitate and ensure consistency in the multiple types of composition that are possible.

Furthermore, System Dynamics practitioners have long recognized and informally utilized higher-level abstractions, such as aging chains, delays, co-flows, target-follower structures, and other ``molecules'' of system structure. While these abstractions are widely used as building blocks, there is no formal mathematical specification of their syntax, structure, or interactions, in a way that would formalize their modular composition with and use within larger models. This gap inhibits the development of reusable, composable models and limits the scalability, flexibility, and clarity of System Dynamics practice.

Additionally, the stratification often employed in System Dynamics models, such as with the use of multiply subscripted model variables and intra-subscript flows to characterize heterogenous complex systems, tends to obscure the underlying dynamics and logic of the model. The bookkeeping associated with such stratification and their entanglement in a single model exhibits poor separation of concerns and often distracts from the understanding of the core system behaviors and can make models less transparent and interpretable.

As a final example, System Dynamics has traditionally privileged Ordinary Differential Equation [ODE]-based interpretations of stock \& flow models.  While key system behavior insights that System Dynamics draws from understanding basic principles of stock \& flow diagrams are not contingent upon treatment of the model as ODEs, ODE-based interpretation of stock \& flow diagrams is virtually universal. While understandable from a historic perspective, this privileging of ODE-based interpretation sacrifices needless potential.  It particularly does so by disregarding the richness and diversity of interpretations that stock \& flow diagrams can support, and which are well-established in the methodological System Dynamics literature.  Alternative interpretations seen in that literature include interpretation of stock \& flow diagrams through as stochastic transition processes, and the lenses of loop gain analysis, multiple types of eigenvalue elasticity analysis, and feedback extraction, to name a few. While tools supporting such alternative lenses for understanding System Dynamics models can and have been built, the failure to provide fully equal footing for a wide diversity of interpretations of stock \& flow models, causal loop diagrams and system structure diagrams constitutes a needless form of semantic inflexibility.  System Dynamics practice and literature testify to the valuable insights that could be offered by routinely supporting a variety of interpretations of System Dynamics diagrams.

This work seeks to informally showcase ways in which such challenges---amongst many others--- can be addressed by the creation a unified mathematical foundation elevating the role of diagrammatic syntax, and enabling flexible, scalable, and stakeholder and mathematically transparent modeling practices.

\subsection{Why a Categorical Approach?}

A categorical approach \cite{SevenSketches,leinster,spivak2014category,mac2013categories} is employed in this work to address the key needs outlined in the previous section, such as the separation of syntax and semantics \cite{MR4605849,baez2022compositional, baez2023categorical}, capturing the relationship between diagrams characterizing a system at varying levels of abstraction \cite{MR4605849, baez2023categorical}, enabling modularity and composition \cite{MR4259613, libkind2020algebra, baez2022compositional}, and enhancing diagrammatic reasoning \cite{baez2023categorical}. These features offer great value in supporting team-based modeling and facilitating systematic comparisons between diagrams -- both within and across distinct types. Such an approach allows for fluid mappings between different diagram types, enabling better learning, reuse, and integration of models.

The concept of composition is central to category theory.  One significant advantage of the approach sketched here is its ability to capture and formalize the composition of models. This framework facilitates the seamless migration of models as new information evolves, separates a model's mathematical structure from its layout, and supports multiscale modeling. For instance, stock \& flow models can be integrated with agent-based characterizations (e.g., alongside state charts) and discrete event components. Furthermore, category theory supports transformations of models for optimization and parallelization, offering substantial performance improvements for large and stratified models. This is particularly relevant for potential alternative semantics involving augmented information and larger amounts of computation, such as particle filtering \cite{li2018applyingMeasles, li2024real, andrieuDoucet2010PMCMC} and Particle Markov Chain Monte Carlo methods \cite{opioids_li2018illuminating,andrieuDoucet2010PMCMC}, where computational efficiency is critical.

The notion of separating of ``syntax'' (the form taken on by some mathematical structure) and ``semantics'' (the interpretation of that structure) is a common theme in category theory.  The foundational elements of the categorical approach described here include encoding form of each System Dynamics diagram ---causal loop diagrams, system structure diagrams, and Stock \& Flow diagrams---as data, subject to well-defined mathematical properties.  More specifically, this is accomplished using the categorical construct of attributed C-sets \cite{patterson-lynch-fairbanks2021}, also known as attributed co-presheaves. The schemas associated with such attributed C-sets can be analogized to a grammar that not only specifies the types of entities present within a diagram---for example, stocks, flows, auxiliary variables--- but also the permissible structure and connectivity between them within a diagram \cite{baez2022compositional, baez2023categorical, MR4605849, Libkind2022, meadows2023hierarchical}. 

This ability to encode a diagram as data subject to rigorous mathematical properties allows for separation of the form of the model from different interpretations of it.  This separation confers flexibility by allowing us to map encoded diagrams to various semantics, including ODE semantics, stochastic state-transition models, and causal loop diagrams. Moreover, the framework allows for advanced analyses, such as eigenvalue elasticity analysis, to be performed on the models.

The concept of \textit{composition} is central to category theory.  A core feature of the categorical approach sketched here is the ability to ``compose'' diagrams---to glue them together---in a variety of ways.  Such composition is carried out via categorical constructs, such as pushouts \cite{MR4483767,fong2015,baezcourser2020} of undirected wiring diagrams \cite{baez2022compositional, baez2023categorical, Libkind2022}, which has the effect of gluing together---or ``indentifying''---common elements in the two models being combined. These constructs enable the visual and logical integration of diagrams into coherent composite models. Additionally, stratification is supported through another piece of categorical machinery, called a pullback \cite{baez2023categorical, Libkind2022}, which allowing for subtle multidimensional layering of model elements without assuming independence across dimensions. 

Application of category theory to an area commonly considers \textit{functors}---translations between mathematical worlds. When applied to System Dynamics models, the categorical approach also enables structure-preserving mappings between diagrams. For example, a system structure diagram---which distinguishes between stocks, flows, and their instantaneous or material connections---can be abstracted into a causal loop diagram with fewer structural distinctions. This abstraction retains key relationships while simplifying the representation.

To allow the benefits of categorical treatment of System Dynamics diagrams to be operationalized in software, this work leverages the algebraic Julia package \texttt{StockFlow.jl} \cite{StockFlow} created by the authors and colleagues, as well as the supporting \texttt{Catlab.jl} platform.  The Julia package StockFlow.jl provides tools for categorically encoding and working with stock \& flow diagrams, causal loop diagrams, and system structure diagrams. Additionally, real-time collaborative modeling is supported through a browser-based interface tool of ModelCollab \cite{modelcollab}, allowing for the manipulation, interpretation, and sharing of diagrams in an accessible manner.  Finally, we also note in this work some innovations introduced by CatColab \cite{CatColab} and its foundation in double categorical theories and models \cite{LAMBERT2024109630}.

\section{Methods}\label{Methods}
\subsection{Basic Concept of Category $\Fin\Set$}
The mathematical foundations of the approach sketched here on the theory of attributed C-sets \cite{patterson-lynch-fairbanks2021}, which provide a formal structure for encoding diagrams and their compositional relationships \cite{li2025enabling}. In this work, we utilize a fundamental concept from category theory that, despite its apparent simplicity, provides a powerful foundation for encoding and manipulating diagrams. This concept involves a category commonly referred to as $\Fin\Set$ -- what would technically be called the skeleton of the category of finite sets. Informally, $\Fin\Set$ contains a collection of objects, each corresponding to a finite set of a specific size. For example, there is one object for a set of size one, another for a set of size two, and so on. Like objects in most categories, those sets in $\Fin\Set$ are linked by arrows (``morphisms").  Each such morphism from set $A$ to set $B$ represents a function from set $A$ to set $B$ -- a map specifying, for each element of set $A$, of a particular element of set $B$ to which it maps.  

\begin{definition}
\textbf{Category} \cite{SevenSketches}: 
A (small) category $\C$ consists of four pieces of data -- objects, morphisms, identities, and a composition rule -- satisfying two properties.

A category $\C$ includes:
\begin{enumerate}
    \item \textit{Objects}: a collection of $Ob(\C)$, elements of which are called objects;
    \item \textit{Morphisms}: for every two objects $c$, $d$, one specifies a set denoted $\C(c,d)$, elements of which are called morphisms (or ``arrows'') from $c$ to $d$. The set of $\C(c,d)$ is often alternatively denoted as $Hom_{\C}(c,d)$, and called the ``hom-set from $c$ to $d$". We can think of this hom-set $\C(c,d)$ as being a set of arrows from $c$ to $d$; it bears emphasis that the interpretation of such an arrow will differ widely between categories;
    \item \textit{Identities}: for every object $c \in Ob(\C)$, a morphism (arrow) $id_c \in \C(c,c)$ called the identity morphism on $c$. $\C(c,c)$ may contain other elements, but only one serves as identity, and its presence is guaranteed;
    \item \textit{Composition} rule: for every three objects $c,d,e \in Ob(\C)$ and morphisms $f \in \C(c,d)$ and $g \in \C(d,e)$, here \textit{must} exist a morphism $f;g \in \C(c,e)$ (or, alternatively, $ g \circ f$) called the composite of $f$ and $g$;   
\end{enumerate}
People sometimes write an object $c \in \C$, instead of $c \in Ob(\C)$. It will also be convenient to denote elements $f \in \C(c,d)$ (i.e., individual morphisms) as $f: c \rightarrow d$. $c$ is called the \textit{domain} of $f$, and $d$ is called the \textit{codomain} of $f$.

These constituents are required to satisfy two conditions that ensure that they ``behave nicely":
\begin{enumerate}
    \item \textit{unitality}: for any morphism $f: c \rightarrow d$, composing with the identity morphisms at $c$ or $d$ on the appropriate side does nothing: $id_c ; f = f$ and $f ; id_d = f$;
    \item \textit{associativity}: for any three morphisms $f: c_0 \rightarrow c_1$, $g: c_1 \rightarrow c_2$, and $h: c_2 \rightarrow c_3$, the following are equal: $(f;g);h=f;(g;h)$. We write this composite as $f;g;h$.
\end{enumerate}
\end{definition}

We normally can use a graph to depict certain categories (without any added constraints), known as \textit{free categories}, where the vertices of the graph represent the objects in the category, and any path in the graph from vertex $x$ to vertex $y$ represents a morphism from $x \rightarrow $ within the category.  In this interpretation, is important to recognize that there is always a 0-length path from vertex $x$ to itself, corresponding to the identity morphism for the corresponding object in the category.

We now introduce the category mentioned above, $Fin\Set$:

\begin{definition}
\textbf{Category of $\Fin\Set$} \cite{SevenSketches}: 
    \begin{enumerate}
        \item Ob($\Fin\Set$) is the collection of all finite sets.
        \item If $S$ and $T$ are sets, then the set of morphisms (the ``hom-set'') between $S$ and $T$ is a set of functions mapping set $S$ into set $T$: $\Set(S,T)={f:S\rightarrow T~|~f \textrm{ is a function}}$.
        \item For each set $S$, the identity morphism is the identity function $id_S:S\rightarrow S$ given by $id_S(s)=s$ for each $s\in S$.
        \item Given functions (morphisms) $f:S\rightarrow T$ and $g:T\rightarrow U$, the composite morphism is the function $f;g:S\rightarrow U$ given by $f;g(s)=g \circ f = g(f(s))$.
    \end{enumerate}
\end{definition}

\subsection{Categorical Representation of Diagrams} \label{Subsection:closedDiagrams}
Although $\Fin\Set$ may initially appear unassuming, it serves as a fundamental building block for mathematically-precise encoding of System Dynamics models as data. Just as stock \& flow diagrams can represent intricate systems, mapping simple structures into $\Fin\Set$ enables the encoding of complex diagrams, such as causal loop diagrams (CLDs), system structure diagrams (SSDs) and Stock \& Flow Diagrams (SFDs), using a consistent, systematic approach that enables use of those representations with a vast range of sophisticated categorical tools -- tools that will allow us to undertake operations such as relating models of the same or different types, model stratification, and composing (``gluing together'' two models into a single composite model).

\subsubsection{Causal Loop Diagrams}

\paragraph{A Causal Loop Diagram as a Functor Mapping from Schema Category to $\Fin\Set$}

\begin{figure}[!h]
\centering
\includegraphics[width=0.5\textwidth]{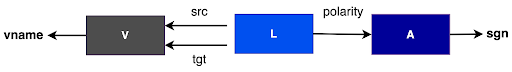}
\caption{The schema category of CLDs ($SchCLDs$).}
\label{figSchCLDs}
\end{figure}

A primitive causal loop diagram (CLD) can be formally encoded as a mapping from a \textit{schema category} -- a category that specifies the ``grammar'' or structure of the diagram, shown in \ref{figSchCLDs} -- to $\Fin\Set$, the category of finite sets. The schema category itself consists of \textbf{objects} linked together by relationships encoded as \textbf{morphisms}. For example:
\begin{itemize}
    \item \textbf{Objects:} Entities such as variables (object \texttt{V}), links (object \texttt{L}), or attributes (object \texttt{A}).  Within  the depiction of the schema category in \ref{figSchCLDs}, such objects are depicted as squares.
    \item \textbf{Morphisms:} Represented as arrows from an object to another object in the schema category, and encode relationships between the types corresponding to such objects. For example, the morphism $src: L \rightarrow V$ characterizes, for a given Link in a causal loop diagram, the source Variable (i.e., the variable from which it originates in that diagram). Similarly, the morphism $tgt: L \rightarrow V$ specifies, for a given link in that diagram, the variable in the diagram that serves as the target of that link.  In a like manner, the morphism $polarity: L \rightarrow A$ allows for mapping a link to its polarity (e.g., positive or negative).
\end{itemize}

Furthermore, \texttt{vname} and \texttt{sgn} are attributes of Variable and Attribute objects, respectively. For example, \texttt{vname} specify, for a given variable, the information of the name of that variable, and \texttt{sgn} maps a given polarities to their signs (``+" or ``-"). 

Schemas characterize the types of entities and their relations.  In order to characterize a given causal loop diagram, we need to instantiate that schema, specifying the data associated with each of the entities and maps between them.  Using a \textit{functor}, the schema category (Figure~\ref{figSchCLDs}) is mapped to $\Fin\Set$. Each such functor can be viewed as an instance of the schema category, which encodes a causal loop diagram:
\begin{itemize}
    \item \textbf{Objects} in the schema category are mapped to objects in $\Fin\Set$ (that, is, to finite sets). 
    \item \textbf{Morphisms} are mapped to morphisms in $\Fin\Set$ (that, is, to functions between these sets).  
\end{itemize}

For instance:
\begin{enumerate}
    \item The object \texttt{V} (variables) maps to a finite set in which each element represents a variable in the diagram.
    \item The object \texttt{L} (links) maps to a finite set in which each element represents a link.
    \item The object \texttt{A} (attributes) maps to a finite set with two elements, $\{+, -\}$, representing positive and negative polarities, respectively.
\end{enumerate}

\textbf{Morphisms} in the schema category are mapped as follows:
\begin{itemize}
    \item The \texttt{src} morphism is mapped to a function specifying, for each link, the variable at which it originates (\texttt{L} $\to$ \texttt{V}).
    \item The \texttt{tgt} morphism is mapped to a function specifying, for each link, its destination variable (\texttt{L} $\to$ \texttt{V}).
    \item The \texttt{polarity} morphism is mapped to a function specifying, for each link, its polarity attribute (\texttt{L} $\to$ \texttt{A}).
\end{itemize}

This functorial encoding preserves the structure of the schema category while representing the diagram concretely in $\Fin\Set$. In a pattern seen across category theory, this mapping from one category to another preserves structure, guaranteeing that the resulting encoding of the diagram is well-behaved.  For a functor, this means that it must map identity morphisms in the schema category to identity morphisms (functions) in $\Fin\Set$, and must map a composition of morphisms within the schema category into a composition of the corresponding functions in $\Fin\Set$.  Such functors from the schema category of CLDs to $\Fin\Set$ provides a formal, systematic approach to representing, manipulating, and analyzing CLDs.

\paragraph{The Category of Causal Loop Diagrams}

One of the notable advantages of the categorical methods underlying the approach outlined here is their ready support for reasoning at a vast range of levels of abstraction, which allows support for a wide range of analyses.  Beyond being amenable to being encoded as a mapping from a schema category into $\Fin\Set$, causal loop diagrams can themselves be organized into a \textbf{category}, with the following constituents:
\begin{itemize}
    \item \textbf{Objects:} Individual CLDs.
    \item \textbf{Morphisms:} Structure-preserving mappings (\textit{homomorphisms}) between CLDs.
\end{itemize}

With causal loop diagrams within the category of causal loop diagrams themselves serving as indivisible objects, this category is independent of any specific way of characterizing or encoding the mathematical structure of such CLDs.  However, recall that a causal loop diagram can itself be described as a functor mapping from the schema category to $\Fin\Set$. Thus, each object in the category of causal loop diagrams can be viewed as constituting such a functor mapping. And each morphism from one such object to another can be regarded as a \textit{natural transformation} between the functor associated with the first such object---itself representing a causal loop diagram---to the second. Just as functors represent structure-preserving mappings between categories, such a natural transformation is a structure-preserving mapping between such functors.  Within this category of causal loop diagrams, structure-preserving mappings (\textit{homomorphisms}) describe how one diagram relates to another. For instance:
\begin{itemize}
    \item Two diagrams with no structural relationship would have no homomorphisms between them.
    \item A causal loop diagram $A$ that is found as a subpiece of another ($B$) would have a distinct homomorphism (morphism within the category of causal loop diagrams) for each such occurrence.
    \item Diagrams that share some structural similarities might have multiple homomorphisms representing different, logically consistent ways in which one diagram can be mapped to another.
\end{itemize}

It bears strong emphasis that while the details of the mathematical details associated with a homomorphism---a structured preserving-mapping---differ between System Dynamics diagrams, any functor from one System Dynamics diagram to another of the same type (e.g., between from Causal Loop Diagram to another, or from one System Structure Diagram to another) captures sensible, intuitive and consistent ways that the first such diagram can be mapped into the second. For example, a given functor might represent a specific way in which one CLD is embedded as a subregion of another -- possibly with additional connections to/from the remainder of the diagram.  Alternatively, a particular functor might characterize \textit{coarse-graining}, in which two or more variables in the first are mapped into a single variable in the latter, and links between variables A and B from the first diagram are mapped on to links of the same polarity between the variables to each of A and B are mapped in the second diagram.  The fact that the mapping represented by the functor is ``structure preserving'' (or that the associated transformation is ``natural'') ensures that the consistency such mappings -- for example, the fact that if two variables are connected by a link from the first to the second of a particular polarity in the first diagram, then they cannot be mapped into variables variables in the second diagram not connected by a link of that polarity and direction. Other such functors can capture mixtures of such mappings.  Yet more general relationships between diagrams can be expressed by a generalization of functor termed a profunctor.

This categorical framework confers several basic advantages:
\begin{itemize}
    \item \textbf{Capturing Formal Relationships Between Diagrams, for example, representing cases where}
    \begin{itemize}
        \item One diagram is \textit{embedded} as a sub-component of another.
        \item One diagram represents a \textit{coarse-graining}---an abstracted, higher-level representation---of another.
    \end{itemize}
    \item \textbf{Diagram Encoding:}
    \begin{itemize}
        \item Each CLD is encoded as a functor from the schema category into $\Fin\Set$, mapping:
        \begin{itemize}
            \item The Variables object (\texttt{V}) to a finite set whose elements represent variables (vertices) in the CLD.
            \item Links object (\texttt{L}) to a finite set whose elements represent links (directional connections) from one variable to another.
            \item Attributes object (\texttt{A}) to a finite set specifying properties of each link, such as its polarity.
        \end{itemize}
    \end{itemize}
    \item \textbf{A Systematic Database Representation \cite{SPIVAK201231}:}
    \begin{itemize}
        \item Within this framework, we can encode the data specified by a functor specifying a System Dynamics diagram in terms of structured tables. Specifically, each object type in the schema category is associated with a structured table. Each row of that table represents each element of the set of items corresponding to that object type (e.g., each stock, or each link) in an instance of a causal loop diagram. And each column corresponds to an outgoing morphism from that object (in database parlance, a foreign key relationship) or an attribute of the object.  Thus, the data associated with a causal loop diagram could be specified by the following tables:
        \begin{itemize}
            \item A \textbf{vertices table} (\texttt{V}) lists variables in the diagram.
            \item A \textbf{links table} (\texttt{L}) identifies connections between variables.
            \item An \textbf{attributes table} (\texttt{A}) specifies properties of links, such as their polarity.
        \end{itemize}
    \end{itemize}
    \item \textbf{Support for Structural Transformations:}
    \begin{itemize}
        \item Relationships such as embedding or coarse-graining are represented as morphisms in the category of CLDs.  As morphisms, they can be themselves be composed, so as to reflect successive structural transformations applied to a CLD, and where each such transformation yields a new CLD.
    \end{itemize}
\end{itemize}

\begin{figure}[!h]
\centering
\includegraphics[width=0.3\textwidth]{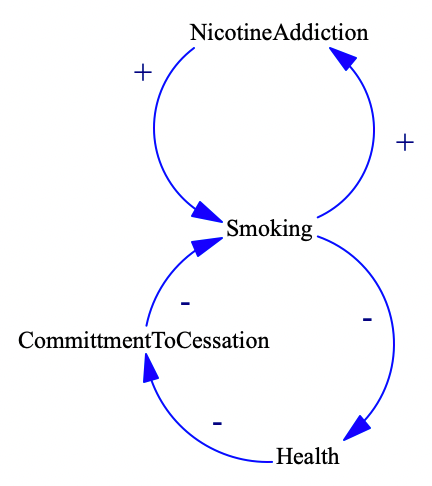}
\caption{An example of a causal loop diagram (CLD) for smoking.}
\label{figexCLD}
\end{figure}

\begin{figure}[!h]
\centering
\includegraphics[width=0.7\textwidth]{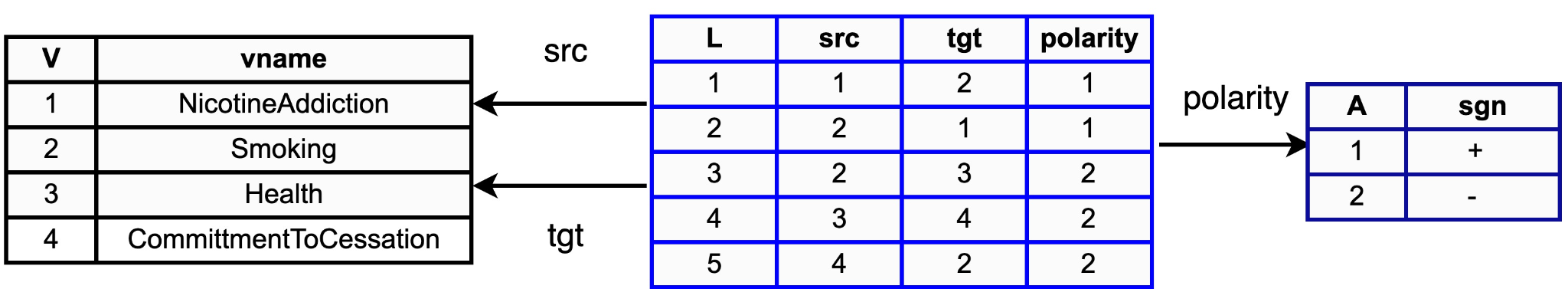}
\caption{Structured tables representing the categorical database for the CLD in Figure~\ref{figexCLD}.}
\label{figexCLDdb}
\end{figure}

\paragraph{Example: Encoding a Causal Loop Diagram}

Figure~\ref{figexCLD} shows a CLD for smoking, involving four variables: ``Nicotine Addiction", ``Smoking", ``Health", ``CommitmenttoCessation". This CLD includes five links, each associated with a polarity. This diagram can be encoded as a specific functor mapping from the schema category (\(SchCLDs\)) to $\Fin\Set$. From a more abstract level, it can also an object of the category of Causal Loop Diagrams. Figure~\ref{figexCLDdb} depicts a categorical database representation of the CLD.

\textbf{Interpretation of Structured Tables:}
\begin{itemize}
    \item \textbf{Variables (\texttt{V})} are encoded by a finite set of four elements, each corresponding to a variable in the CLD. Names are encoded in the \texttt{vname} column.
    \item \textbf{Links (\texttt{L})} are encoded by a finite set of five elements, each corresponding to a link. Columns \texttt{src}, \texttt{tgt}, and \texttt{polarity} (the outgoing morphisms from object (\texttt{L}) in the schema category $SchCLDs$) encode the source, target, and polarity of links, respectively. 
    \item \textbf{Attributes (\texttt{A})} are encoded by a finite set of two elements (\(\{+, -\}\)), each serving as one of the possible polarities of links in the \texttt{sgn} column.
\end{itemize}

This categorical framework provides a rigorous mathematical foundation for encoding, comparing, and manipulating CLDs. It supports scalable and systematic reasoning about the structured relationships between system models, and transformations that can be performed on them while preserving their essential structure.

\subsubsection{System Structure Diagrams}\label{SSDPolarityReasoning}

\begin{figure}[!h]
\centering
\includegraphics[width=0.8\textwidth]{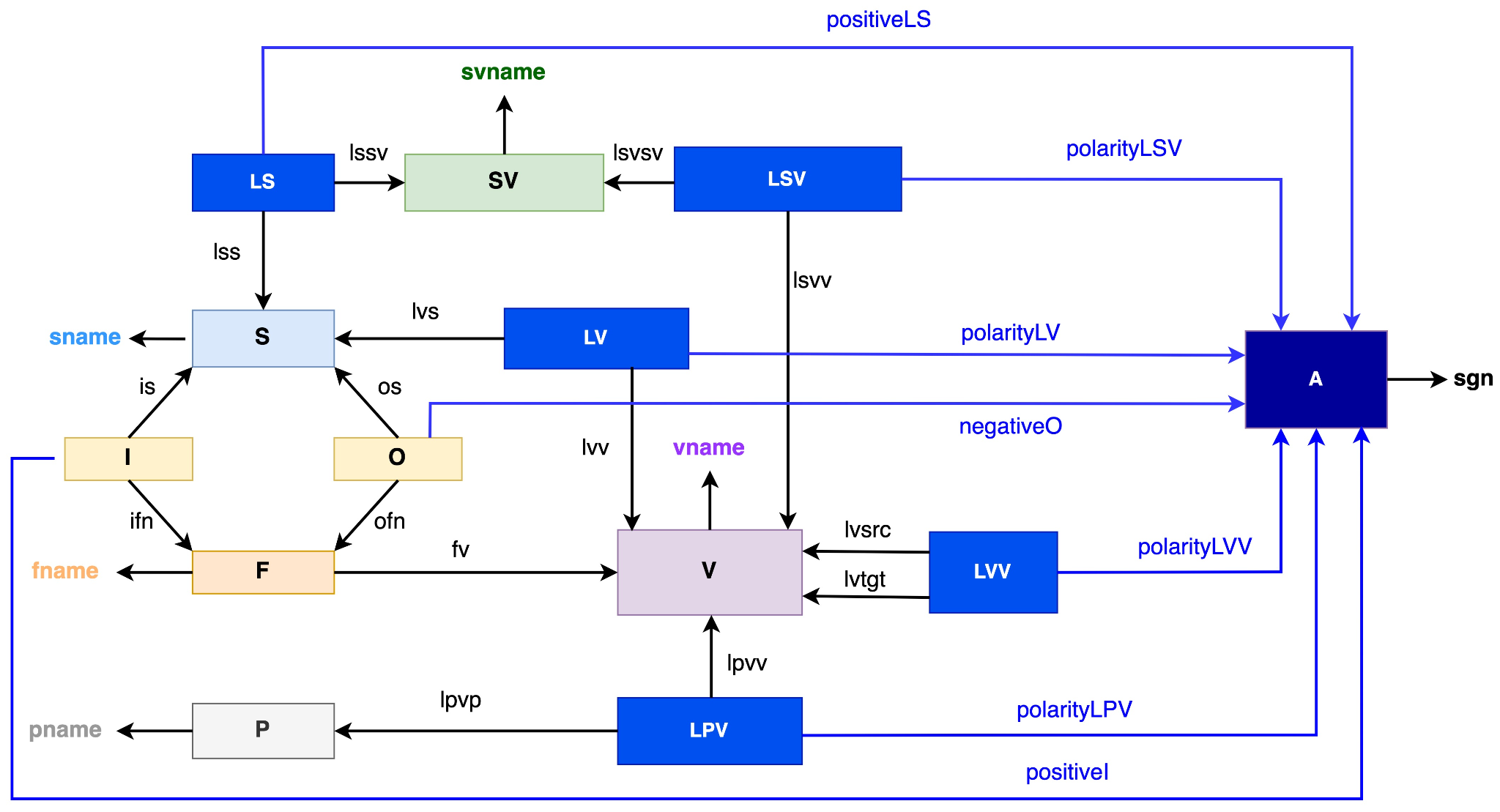}
\caption{The schema category of SSDs ($SchSSDs$).}
\label{figSchSSDs}
\end{figure}

\paragraph{A Categorical Characterization of System Structure Diagrams}
System structure diagrams (SSD) can be described according to a similar categorical framework as causal loop diagrams, but with a more involved schema. As readers of System Dynamics Review will be aware, these diagrams capture additional distinctions and sharpen the representation of a system by distinguishing stocks, flows, auxiliary variable, and the relationships between them. Consequently, the schema category for system structure diagrams---shown in Figure~\ref{figSchSSDs} and labeled as $SchSSDs$---is significantly richer and more textured than that for causal loop diagrams.

The interpretation of the schema category of SSDs involves the following:
\begin{itemize}
    \item \textbf{Objects in the Schema Category:} The SSD schema category includes objects representing the following:
    \begin{itemize}
        \item Stocks (\texttt{S}), to allow for characterizing stocks in SSDs.
        \item Flows (\texttt{F}), to support characterization of flows in SSDs.
        \item Inflows (\texttt{I}), to support representation of the inflows of stocks in SSDs.
        \item Outflows (\texttt{O}), to enable characterization of the outflows of stocks in SSDs.
        \item Auxiliary Variables (\texttt{V}), similarly indicating components of auxiliary variables in SSDs.
        \item Sum Auxiliary Variables (\texttt{SV}), indicating components of a specific type of auxiliary variables, where the sum auxiliary variables is implicitly associated with a formula consisting of the sum of all stocks linked to them.
        \item Parameters (\texttt{P}), to allow for encoding (constant) parameters in SSDs, which serve as a means of specifying assumptions regarding exogenous quantities.
        \item Attributes (\texttt{A}) indicates the attributes of links, such as polarities (positive or negative).  This object operates in a manner similar to that for the attribute object (\texttt{A}) in the schema category of CLDs (Figure~\ref{figSchCLDs}).
        \item Links (\texttt{LS}, \texttt{LSV}, \texttt{LV}, \texttt{LVV}, \texttt{LPV}) support encoding of dependencies of the objects to which those links connect, such as those between stocks and auxiliary variables (\texttt{LV}), those between auxiliary variables (\texttt{LVV}), those between stocks and sum auxiliary variables (\texttt{LS}), between sum auxiliary variables and auxiliary variables (\texttt{LSV}), and those between constant parameters and auxiliary variables (\texttt{LPV}).
    \end{itemize}
    \item \textbf{Morphisms in the Schema Category:} Morphisms between objects in the schema category are designed to support encoding the relationships between objects:
    \begin{itemize}
        \item Morphisms \texttt{is} and \texttt{ifn}, support representation of the inflow \texttt{I} relationships between stocks \texttt{S} and flows \texttt{F}. Similarly, morphisms \texttt{os} and \texttt{ofn} enable representation of the outflow relationships.
        \item Morphism \texttt{fv} enables the assignment of an auxiliary variable for each flow, where that auxiliary variable carries the formula corresponding to the flow.  This structure reflects a simplification associated with the representation of stock and flow diagrams herein, whereby we forgo the extra structure required to directly associate formulas with flows, and instead simply associate a (hidden) auxiliary variable with each flow.
        \item The pair of outgoing morphisms for link objects -- the two black arrows associated with each link object in the free graph representing the schema category shown in Figure~\ref{figSchSSDs} -- define the source and target of each link. For example, for the link type \texttt{LV}, the outgoing morphism \texttt{lvs} represents the source of the link type \texttt{LV} originating from a stock, while the outgoing morphism \texttt{lvv} represents the target of the link type \texttt{LV} pointing to an auxiliary variable. Other link types, such as \texttt{LS}, \texttt{LSV}, \texttt{LVV}, and \texttt{LPV}, similarly have a pair of outgoing morphisms that specify both the source and the target.
        \item The outgoing morphisms -- represented by the blue arrows -- indicate the polarities of links (\texttt{LS}, \texttt{LSV}, \texttt{LV}, \texttt{LVV}, \texttt{LPV}), as well as inflows (\texttt{I}) and outflows (\texttt{O}). The polarities of these objects are as follows:
        \begin{itemize}
            \item \texttt{LS} maps to positive polarity.  This reflects the fact that an increase in each of the variables on which a sum (specifically) auxiliary variable depends serves to increase the value of that auxiliary variable.
            \item  \texttt{I} also maps to positive polarity.  This reflects the fact that an increase in the inflow to a stock serves to increase the value of that stock.
            \item \texttt{O}, by contrast, always maps to a negative polarity, reflecting the fact that an increase in an outflow tends to reduce the value of the upstream stock.
            \item The polarities of all other links (\texttt{LSV}, \texttt{LV}, \texttt{LVV}, and \texttt{LPV}) pointing to auxiliary variables may map to either positive or negative polarity.
        \end{itemize}
    \end{itemize}
\end{itemize}

Similar to the approach applied for encoding Causal Loop Diagrams, a System Structure Diagram (SSD) is represented as a functor mapping from the schema category ($SchSSDs$, shown in Figure~\ref{figSchSSDs}) of SSDs to the category $\Fin\Set$. This functor maps each object (e.g., \texttt{S}) in the schema category to a finite set containing corresponding elements in the SSD (for \texttt{S}, the stocks in the diagram).  Similarly, the functor maps each morphism (e.g., that from \texttt{I} to \texttt{S}) to a function representing the relationships between the corresponding sets (here, specifying, for each inflow, the stock into which it flows). 

While the preceding discussion has focused on the characterization of a System Structure Diagrams, as for causal loop diagrams, the underlying mathematics gives rise to structure at a higher level of abstraction --- namely, a category of SSDs in which each object represents a System Structure Diagram, and each morphism represents a structure-preserving mapping (\textit{homomorphisms}) between two SSD objects. A homomorphism from one SSD to another allows for characterizing relationship between such diagrams.  For example, a specific SSD homomorphism $A \rightarrow B$ might characterize that SSD $A$ is embedded in SSD $B$, or is coarse-grained in a particular way within diagram $B$, or some mixture thereof.

Finally, like their causal loop diagram counterparts, each System Structure Diagram can be viewed as encoded in a categorical database, which can serve as a data structure to systematically represent the model data in a way that captures its mathematical structure and is suitable for model analysis and transformation.

\begin{figure}[!h]
\centering
\includegraphics[width=0.8\textwidth]{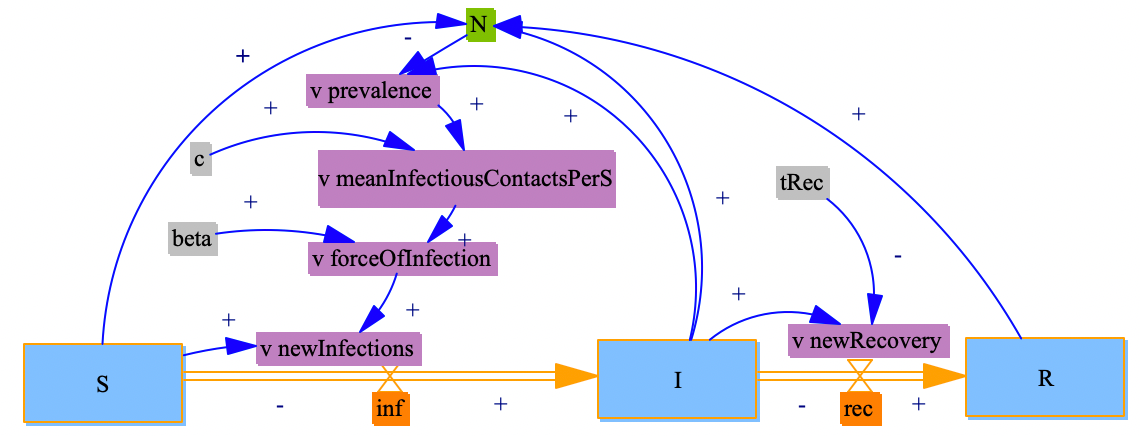}
\caption{An example of a System Structure Diagram.}
\label{figexSSD}
\end{figure}

\begin{figure}[!h]
\centering
\includegraphics[width=0.9\textwidth]{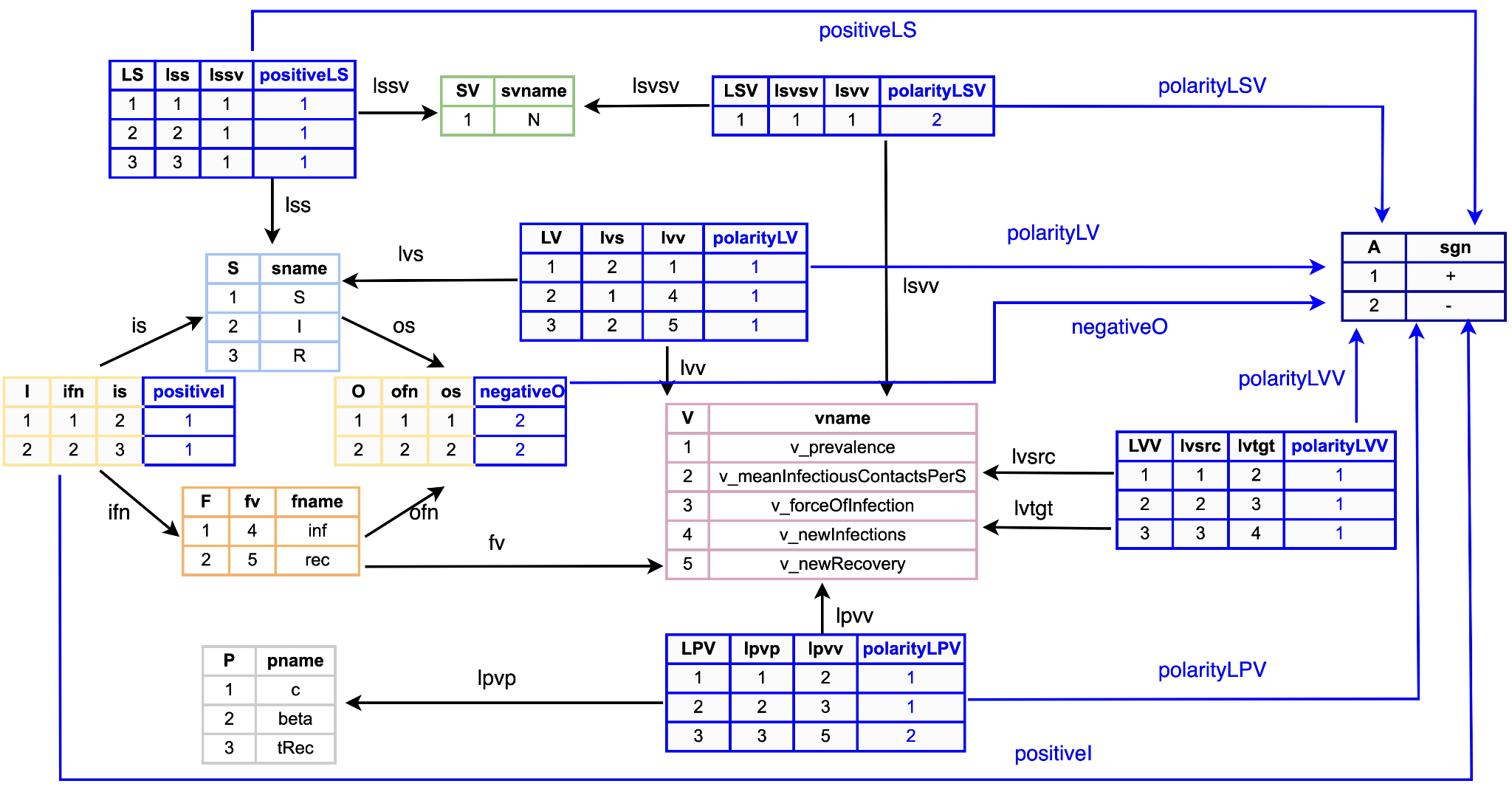}
\caption{The categorical database for the System Structure Diagram in Figure~\ref{figexSSD}.}
\label{figexSSDdb}
\end{figure}

\paragraph{Example: Encoding a System Structure Diagram}

Figure~\ref{figexSSD} illustrates a System Structure Diagram (SSD) for an infectious disease SIR (Susceptible-Infective-Recovered) model \cite{sterman2000business}. Using the framework presented in this work, this SSD is encoded as a functor mapping from the SSD schema category (\(SchSSDs\)) to $\Fin\Set$.  Figure~\ref{figexSSDdb} presents the categorical database representation of the SIR SSD depicted in Figure~\ref{figexSSD}. 

The following represents the interpretation of the encoded structured tables representing the SIR System Structure Diagram:

\begin{itemize}
    \item \textbf{Stocks (\texttt{S})} consist of a finite set of three stocks -- ``S", ``I", and ``R". Names are encoded in the attribute column of \texttt{sname}.
    
    \item \textbf{Auxiliary Variables (\texttt{V})} are given by a finite set of five auxiliary variables -- ``v\_prevalence", ``v\_meanInfectiousContactsPerS", ``v\_forceOfInfection", ``v\_newInfections", and ``v\_newRecovery". Names are encoded in the attribute column \texttt{vname}.
    
    \item \textbf{Flows (\texttt{F})} are given by a finite set of two flows -- ``inf" and ``rec". The column \texttt{fv} (the outgoing morphism of object \texttt{F} in the schema category) assigns each flow a formula represented by an auxiliary variable. For example, the value of flow ``inf" corresponds to the auxiliary variable ``v\_newInfections". Similarly, the value of flow ``rec" corresponds to the auxiliary variable ``v\_newRecovery". Names are encoded in the attribute column of \texttt{fname}.
    
    \item \textbf{Parameters (\texttt{P})} are represented by a finite set of three constant parameters -- ``c", ``beta", and ``tRec". Names are encoded in the attribute column \texttt{pname}.
    
    \item \textbf{Sum Auxiliary Variables (\texttt{SV})} are specified by a finite set of one sum auxiliary variable -- ``N". Names are encoded in the attribute column \texttt{svname}.
    
    \item \textbf{Attributes (\texttt{A})} are given by a finite set of two elements (\(\{+, -\}\)), encoding the polarity of links in the \texttt{sgn} column.
    
    \item \textbf{Inflows (\texttt{I})} correspond to a finite set of two elements of inflows. Columns \texttt{ifn} and \texttt{is} specify which flow serves as an inflow to which stock. For example, the first row of the table \texttt{I} indicates that the flow ``inf" (with index 1 in the flow table \texttt{F}) is an inflow to the stock ``I" (with index 2 in the stock table \texttt{S}). The column \texttt{positiveI} always has a value of ``1" for each row, indicating the positive polarity of inflows.
    
    \item \textbf{Outflows (\texttt{O})} are given by a finite set of two elements, each an outflow in the diagram. The interpretation of the outflow table (\texttt{O}) is similar to that for the inflow table (\texttt{I}). The main difference is that the outflow table specifies the outflow (rather than the inflow) relationships between flows and stocks. The column \texttt{negativeO} always has a value of ``2" for each row, indicating the negative polarity of outflows.
    
    \item \textbf{Links from Stocks to Sum Auxiliary Variables (\texttt{LS})} correspond to a finite set of three links from the source stocks (column \texttt{lss}) -- ``S", ``I", and ``R" (indices 1, 2, and 3, respectively, in the stock table \texttt{S}) -- to the target sum auxiliary variable ``N" (total population), which carries an index of 1 in the table of sum auxiliary variables \texttt{SV}. The column \texttt{positiveLS} always has a value of ``1", indicating positive polarity.
    
    \item \textbf{Links from Stocks (\texttt{LV}), Parameters (\texttt{LPV}), Sum Auxiliary Variables (\texttt{LSV}), Auxiliary Variables (\texttt{LVV}) to Auxiliary Variables} are given by four tables, each representing a finite set of links from a specific type of source to the target auxiliary variables. For example, in the links table from stocks (\texttt{LV}), columns \texttt{lvs} and \texttt{lvv} specify links from stocks (e.g., stock index 2 of ``I" in the first row) to auxiliary variables (e.g., auxiliary variable index 1 of ``v\_prevalence" in the first row). The column \texttt{polarityLV} indicates the polarity of each link (either positive or negative). For instance, the polarity value ``1" in the first row of table \texttt{LV} signifies a positive relationship between stock ``I" and auxiliary variable ``v\_prevalence". The other three link tables (\texttt{LPV}, \texttt{LSV}, and \texttt{LVV}) follow a similar structure.
\end{itemize}

The SSD diagram encoded by this categorical database itself corresponds to an object in the category of System Structure Diagrams, with morphisms to other such objects (SSDs) given by homomorphisms.


\subsubsection{Stock \& Flow Diagrams}
\paragraph{Categorical Encoding of Stock \& Flow Diagrams}
The categorical mathematics used to encode the structure of causal loop diagrams and system structure diagrams can also be applied for stock \& flow diagrams.  In a manner similar to the other diagram types, stock \& flow diagrams form a category in which each object corresponds to a specific stock \& flow diagram, and morphisms correspond to structure-preserving mappings (homomorphisms) between such diagrams. 

\begin{figure}[!h]
\centering
\includegraphics[width=0.6\textwidth]{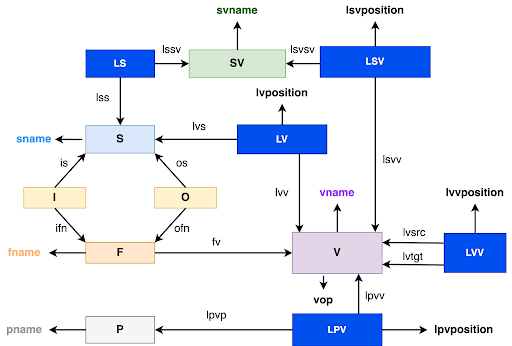}
\caption{The schema category of stock \& flow diagrams ($SchSFDs$).}
\label{figSchSFDs}
\end{figure}

Figure~\ref{figSchSFDs} shows the schema category (labeled as $SchSFDs$) for the structure of stock \& flow diagrams.

\begin{figure}[!h]
\centering
\includegraphics[width=0.8\textwidth]{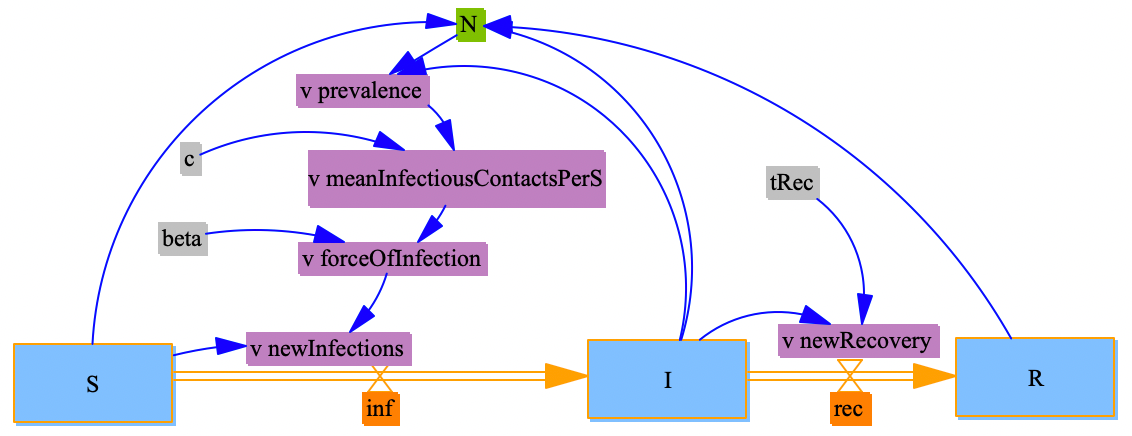}
\caption{An example of a stock \& flow diagram.}
\label{figexSFD}
\end{figure}

The schema category for SFDs ($SchSFDs$) is very similar to the schema category for SSDs ($SchSSDs$). By comparing the schema category $SchSFDs$ in Figure~\ref{figexSFD} with $SchSSDs$ in Figure~\ref{figexSSD}, the schema category $SchSFDs$ exhibits the following differences:
\begin{itemize}
    \item It excludes both the attribute object $A$ and the morphisms (blue arrows) pointing to object $A$, which carry the information about the polarities of links, inflows, and outflows. This is because SFDs do not in general explicitly characterize such polarities.
    \item It includes five additional attributes that provide the information required to support encoding of the formulas for auxiliary variables/flows in SFDs as abstract syntax trees:
    \begin{itemize}
        \item The attribute \texttt{vop} of object \texttt{V} carries the information about the operator associated with the whole or component of a formula, as it would be used in each auxiliary variable.
        \item The attributes \texttt{lvposition} of object \texttt{LV}, \texttt{lsvposition} of object \texttt{LSV}, \texttt{lvvposition} of object \texttt{LVV}, and \texttt{lpvposition} of object \texttt{LPV}, carry positional information for the source object of the corresponding types of links in the formulas.  For example, for an operator ``*'', a value of 2 for \texttt{lvposition} would indicate that the corresponding stock consists of the second argument to the multiplication operator.
    \end{itemize}
\end{itemize}

Similar to CLDs and SSDs, stock \& flow diagrams (SFDs) are also encoded using a database-like format, where the elements of the diagram --- such as stocks, flows, auxiliary variables, and their relationships --- are encoded as data. As a key component of this database representation, the formulas associated with dynamic or auxiliary variables, including flows, are explicitly encoded as data. 

We now turn to an example illustrating this categorical database structure and encoding process.

\paragraph{Example: Encoding a Stock \& Flow Diagram}

\begin{figure}[!h]
\centering
\includegraphics[width=0.9\textwidth]{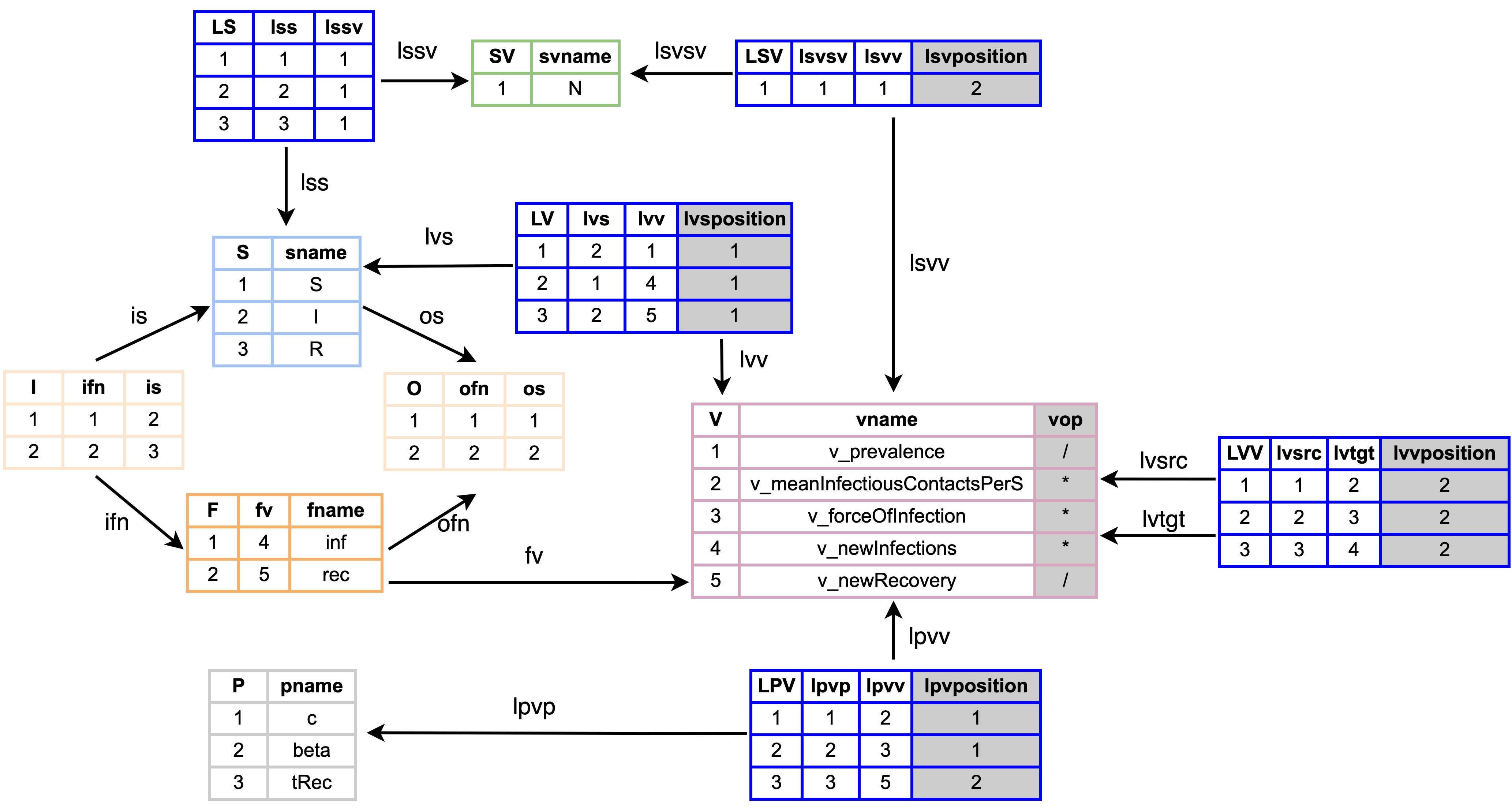}
\caption{The encoded model data for the example stock \& flow diagram in Figure~\ref{figexSFD}, as represented via a categorical database.}
\label{figexSFDdb}
\end{figure}

Figure~\ref{figexSFDdb} illustrates the categorical database structure used to encode a SFD representing the same infectious disease SIR (Susceptible‐Infective‐Recovered) context as was employed for the SSD example, and which is depicted in \ref{figexSFD}. The figure depicts this SIR stock \& flow diagram. Because the schema category of SFDs exhibits a great deal of common substructure with the schema category of SSDs, we will not provide a detailed interpretation of each table and column in the categorical database shown in Figure~\ref{figexSFDdb}. 

However, it is important to note a significant difference between stock \& flow diagrams (on the one hand) and both causal loop diagrams and system structure diagrams (on the other). Stock \& flow diagrams can quantitatively represent a mathematical model (most commonly, ordinary differential equations, ODEs) at a sufficiently granular level that it precisely characterizes a dynamical system.  While both causal loop diagrams and system structure diagrams provide a great deal of information concerning the structure underlying a dynamical system, that information is insufficient to unambiguously characterize system evolution. 

Given this distinction, we will focus here on distinctive features of SFDs, in the form of algorithms for encoding the formulas of auxiliary variables and mappings to interpret such diagrams according to the most common semantics--- ordinary differential equations (ODEs)---using the data encoded regarding such diagrams.  It is to be emphasized that while ODEs constitute a widespread interpretation of stock \& flow diagrams, they do not constitute a privileged semantics, and Section \ref{sec:OtherSemantics} discusses alternative interpretations of such diagrams.

\paragraph{Algorithms for Generating Formulas for Auxiliary Variables in the SFDs}
The following describes the derivation of formulas for ``Sum Auxiliary Variables (SV)" and ``Auxiliary Variables (V)" using the categorically-encoded data.

\begin{itemize}
    \item \textbf{Sum Auxiliary Variables (\texttt{SV}):} The formulas for "Sum Auxiliary Variables (\texttt{SV})" are defined as the sum of all stocks to which they are linked. For the example in Figure~\ref{figexSFD} and Figure~\ref{figexSFDdb}, the mathematical equation for the sum auxiliary variable \(N\) is represented as \(N = S + I + R\), where \(S\), \(I\), and \(R\) are the stocks offering connections to \(N\).

    \item \textbf{Auxiliary Variables (\texttt{V}):} The formulas for ``Auxiliary Variables (\texttt{V})" --- including those associated with flows --- are derived based on the following rules:
    \begin{itemize}
        \item \textbf{Operators:} The operators applied in the formula are represented by the attributes of the auxiliary variable \texttt{V}, as given by the column of ``vop" in the encoded database. 
        \item \textbf{Arguments:} For a given auxiliary variable (as delineated by the row within the ``V'' table), the terms within the formula to which the operator is applied---that is, the arguments to that operator---are drawn from the source of any link among \texttt{LSV}, \texttt{LV}, \texttt{LVV}, and \texttt{LPV}, whose target is this variable. The positions of the arguments in the formulas are represented by the ``position'' attributes of those link tables. 
        
        \item \textbf{An Example:} The operator of the auxiliary variable ``v\_prevalence" is division (\texttt{/}), as indicated by the first row of the table \texttt{V} in Figure~\ref{figexSFDdb}). ``I'' is an argument in the formula of the auxiliary variable ``v\_prevalence" (derived from the first row of the table \texttt{LV} in Figure~\ref{figexSFDdb}), and the position of ``I'' in the formula is ``1'', which indicates it serves as the \texttt{numerator} of the operator \texttt{/}. ``N" is another argument (derived from the only row of the table \texttt{LSV} in Figure~\ref{figexSFDdb}), and the position (the column of ``lsvposition") of ``N" is ``2", which indicates ``N" is the \texttt{denominator} of the operator \texttt{/}. Thus, the formula of the auxiliary variable ``v\_prevalence" is $I/N$.
    \end{itemize}
\end{itemize}

\subsubsection{Summary of Methodological Benefits}

System Dynamics diagrams have long been encoded and manipulated by purpose-built software, from Dynamo \cite{Dynamo} to STELLA/Ithink \cite{Stella}, Vensim \cite{Vensim}, Powersim \cite{Powersim}, and beyond.  While such software can manipulate System Dynamics diagrams in powerful ways, the diagrams within such software are encoded using data structures of general purpose programming languages, and manipulated using supporting algorithms.  While such mechanisms are widely used and efficient, they fail to make explicit the underlying mathematical relationships.  Such mathematical opacity greatly limits flexibility.  

By contrast, encoding diagrams as structured data joined by mathematically well-defined relationships using the approach sketched here enables several powerful capabilities:
\begin{itemize}
    \item \textbf{Reasoning and Transformation:} Mathematical encoding allows for reasoning about the diagrams, safely transforming them, and ensures structural consistency.
    \item \textbf{Analysis and Mapping:} Diagrams can be analyzed rigorously, mapped to different semantics, and related to other diagrams within the same category.
    \item \textbf{Integration Across Models:} Encoded diagrams can be linked, compared, or integrated with other models to explore System Dynamics in larger contexts.
    \item \textbf{Integration Across Levels of Abstraction:} An instance of one System Dynamics diagram can be systematically mapped to and compared against diagrams at higher levels of abstraction.

    \item \textbf{Modular Software Support:} Given careful design, software designed to understand the mathematical characterizations sketched here \cite{patterson-lynch-fairbanks2021} allow for modular extension to extend its functionality -- for example, by supporting additional types of comparisons between diagrams of a particular types, additional types of composition, mappings to additional semantic domains, and diagram transformations.
\end{itemize}

By encoding these three System Dynamics diagram types categorically, this approach facilitates systematic reasoning, rigorous analysis, and the ability to map between different representations of a model. This categorical representation thus provides a robust mathematical foundation for working with those diagrams in a consistent and scalable way.

\subsection{A Composition Framework}\label{Compose}

A central goal of this framework is to enable the construction of larger, integrated diagrams from smaller, modular components. These smaller diagrams may represent subsystems.  For a model of West Nile Virus \cite{DecisionTreeWNV}, of example, subsystems might characterize 1) the aquatic (egg, larval and pupal phases of mosquito lifecycle) 2) adult phases of mosquito lifecourse 3) an avian reservoir population and 4) infection among humans as dead-end hosts.  Alternatively, within a human communicable disease transmission model subsystems could include 1) the vaccination process and 2) natural history of infection 3) acute care processes associated with severe and critical infections.  Alternatively, the smaller diagrams could represent higher-levels of patterns in System Dynamics practice, such as ``molecules" \cite{hines1996molecules} like first-order delays, coflows, and aging chains. The ability to compose such diagrams is of great value in building scalable, flexible, extensible models of complex systems, and for testing, debugging, sharing and reusing model components.

\begin{figure}[!h]
\centering
\includegraphics[width=0.5\textwidth]{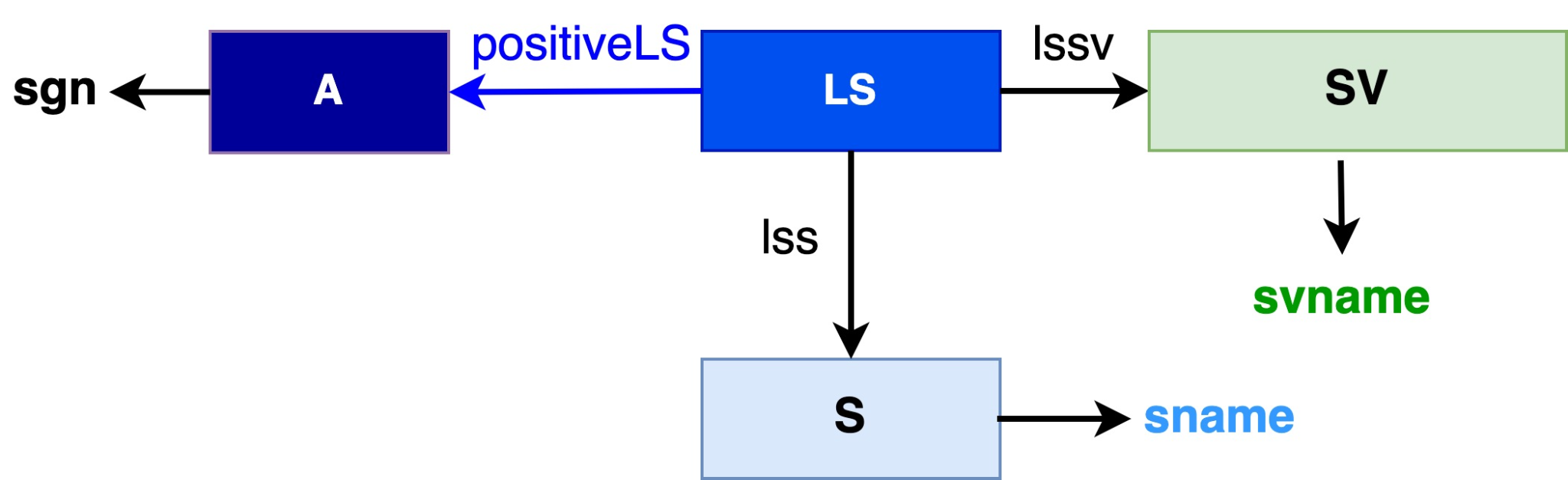}
\caption{The interface schema category for composing System Structure Diagrams.}
\label{figSchSSDsFeet}
\end{figure}

\begin{figure}[!h]
\centering
\includegraphics[width=0.3\textwidth]{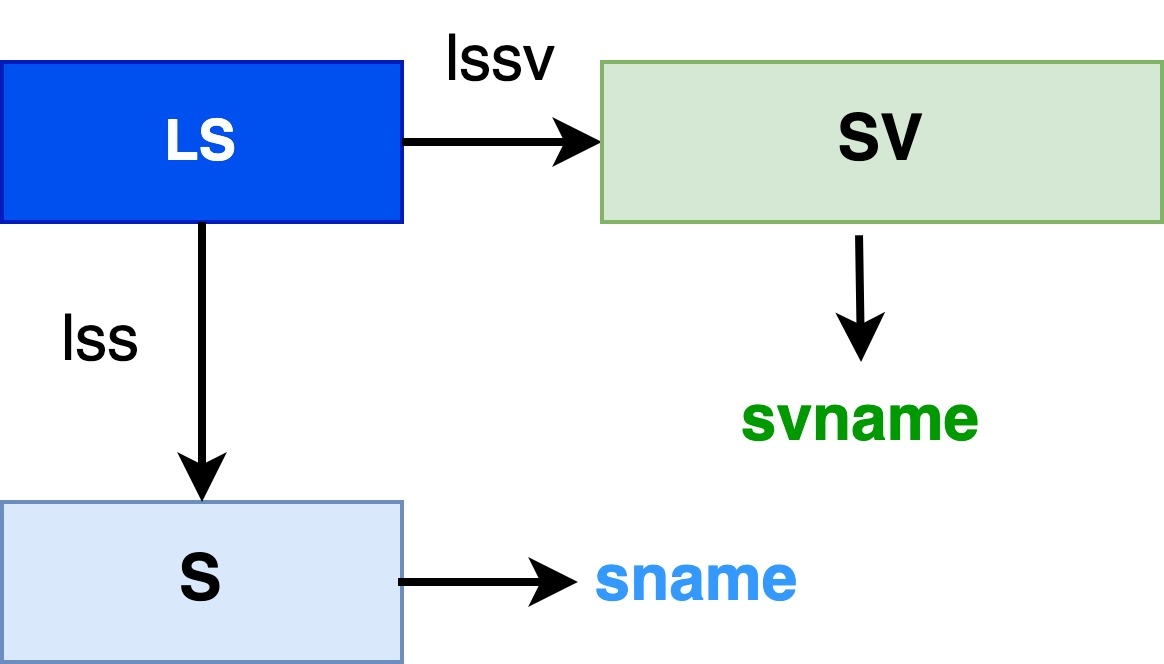}
\caption{The interface schema category for composing stock \& flow diagrams.}
\label{figSchSFDsFeet}
\end{figure}

To achieve the ability to compose diagrams, we utilize the mathematical mechanism of structured/decorated cospans \cite{fong2015, MR4483767, baezcourser2020}. By leveraging these structures, we can construct open diagrams, which are diagrams with well-defined interfaces. By composition, we refer to the ability to ``stick together'' two or more open diagrams to create a larger diagram by identifying shared interfaces as points in common between the constituent diagrams. Thus, in composing two diagrams, certain substructure is treated as being shared between those diagrams, and thus serves to ``glue together" such diagrams into a larger whole. The pairing of such diagrams with one or more suitable interfaces for possible gluing gives rise to an ``open'' version of that diagram.  Appropriate design of corresponding interfaces allows for application of this approach to any of the three types of System Dynamics diagrams covered here, allowing, in turn, for the definition of \textit{open} causal loop diagrams, \textit{open} system structure diagrams, and \textit{open} stock \& flow diagrams, .

To perform the actual gluing of these diagrams at specified interfaces, we make use of a key concept from category theory known as the \textit{pushout} \cite{SevenSketches}. Taking the pushout of two open diagrams together with a specific interface in common---an interface that delineates which part of each diagram is to be ``glued'' to which part of the other diagram---allows us to combine open diagrams along that interface in a mathematically consistent manner.

In the following sections, we will introduce the mathematical structures underlying this approach, explaining how structured/decorated cospans enable the construction of open diagrams, and how their composition is performed by the pushout operation.

\subsubsection{A Categorical Structure of Open Diagrams}
\paragraph{Mathematical Basis of Open Diagrams}

\begin{figure}[!h]
\centering
\includegraphics[width=0.2\textwidth]{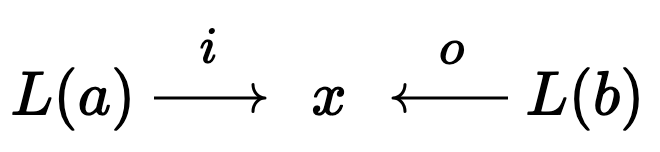}
\caption{The mathematical construction of structured cospan.}
\label{figstrcospan}
\end{figure}

A key feature of this framework is the separation of a system’s internal structure from its interface. In alignment with concepts from software engineering, this separation defines ``interfaces" or ``ports" through which diagrams can connect. The internal structure of a system is represented by one category (denoted as category $\X$), while its ports are described in a simpler category (denoted as category $\A$) that defines the connection points. For example, in stock \& flow diagrams, ``ports" may correspond to collections of stocks, links, or sum auxiliary variables (auxiliary variables that summarize stocks). In causal loop diagrams, ports are defined around variables and links, ensuring that polarities are shared during composition.

Figure \ref{figstrcospan} depicts the mathematical structure of a structured cospan.

Let $\mathcal{X}$ denote the relevant category of diagrams. For example, depending on interests, $\mathcal{X}$ may represent the category of causal loop diagrams, the category of system structure diagrams, or the category of stock \& flow diagrams (introduced in Section \ref{Subsection:closedDiagrams}). An object $x$ in category $\X$ represents an instance of a specific diagram, such as a causal loop diagram, a system structure diagram, or a stock \& flow diagram.  This diagram will represent the model which we which to equip with interfaces to support composition with other models.

The next task is to define the interfaces to that model.  To do so, we define a schema category that captures the structure of interfaces specific to the corresponding diagram type (e.g., causal loop diagrams). In general, this may include not just objects---representing components such as vertices and links---but also relationships between them. Figure~\ref{figSchSSDsFeet} illustrates the schema category of the interface for to allow a system structure diagram to interface to other SSDs.  By contrast, Figure~\ref{figSchSFDsFeet} presents the schema category of the composition interface for stock \& flow diagrams. In application, the composition interfaces are typically subparts of the full diagrams. Consequently, the schema categories of the composition interfaces for both system structure diagrams (Figure~\ref{figSchSSDsFeet}) and stock \& flow diagrams (Figure~\ref{figSchSFDsFeet}) are subparts of the schema categories of the corresponding type of diagrams, as shown in Figure~\ref{figSchSSDs} and Figure~\ref{figSchSFDs}, respectively. However, in the application of composing causal loop diagrams, we can enable the composition by identifying any subpiece of a causal loop diagram.  That is, if we wish to compose causal loop diagram A with another such diagram B, we can do so not only by identifying certain vertices in CLD A with corresponding vertices of CLD B, but can also identify a link of a given polarity link in CLD A with a corresponding link of identical polarity in CLD B. It follows that the schema category for the interface for causal loop diagrams is identical to the schema category of the causal loop diagrams themselves, as depicted in Figure~\ref{figSchCLDs}.

Mathematically, we can construct a category of interfaces from the schema category of interfaces for the diagram type of interest.  Any possible such interface is given by an instance of that schema category -- that is, by an attributed C-Set, a (functor) mapping the schema category of the interface to the category $\Fin\Set$. We denote as category $\A$ the category in which each object is such a possible interface, and morphisms are structure-preserving transformations (homomorphisms) between them.  In Figure~\ref{figstrcospan}, $a$ and $b$ are two objects in this category $\A$ that are designated to serve as ports around which to compose diagrams; it bears emphasis that for all of the uses considered in this paper, each of $a$ and $b$ will be an instance of the schema for interfaces for the corresponding diagram type -- that is, an attributed C-Set, where ``C'' is that schema.  However, at this point, reflecting the shift to a higher level of abstraction, it is best to think of each of them simply as an object in the category $\A$.  A functor $L: \A \to \X$ establishes the mapping from the interface diagram category $\A$ to the category $\mathcal{X}$ associated with the appropriate type of diagram.  Because this functor can map an object in $\A$ to an object in $\X$, and similarly for morphisms, it serves to map $a$ to a corresponding object in $\X$ (for our case, thus promoting an instance for the schema in the foot to an instance for the schema for the full diagram, in the apex).  Doing so for both feet $a$ and $b$ and specifying the corresponding structure-preserving mappings (homomorphisms) on the legs of the cospan (mappings from the ``promoted'' foot to the instance in the apex) serves to construct the structured cospan.  For a given leg, such a homomorphism could be regarded as picking out which items in the interface correspond to which items in the diagram in the apex of the cospan. This structure is used for open version of each type of System Dynamics diagram, including open stock \& flow diagrams, open system structure diagrams, and open causal loop diagrams.

\paragraph{Examples of Open Diagrams}

\begin{figure}[!h]
\centering
\includegraphics[width=0.8\textwidth]{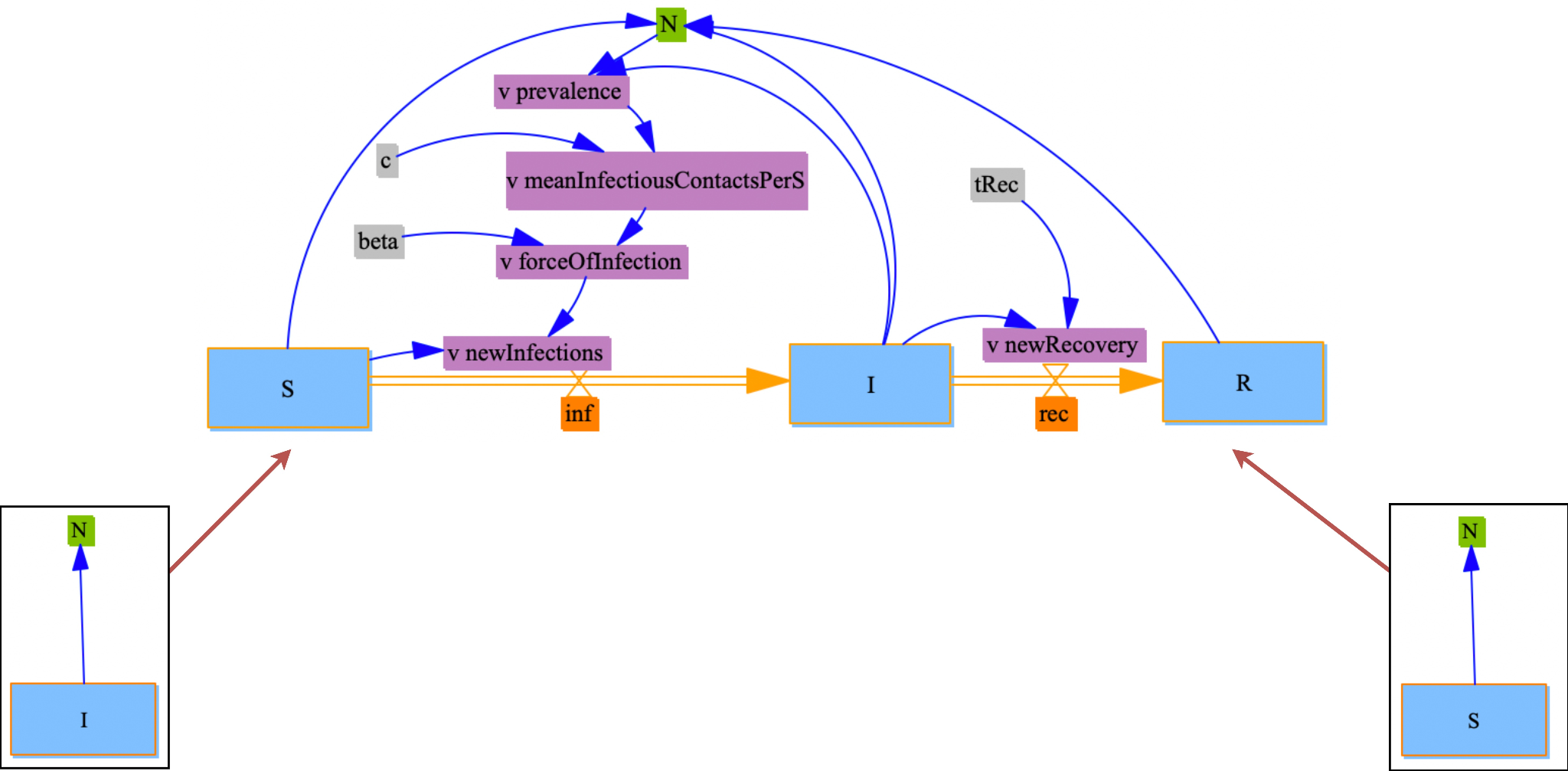}
\caption{An example of an open stock \& flow diagram.}
\label{figexopensfd}
\end{figure}

Figure~\ref{figexopensfd} shows an example of an open stock \& flow diagram featuring an SIR model.  As shown in the figure, there is an object $x$ at the apex of the structured cospan, as characterized in Figure~\ref{figstrcospan}, with two interfaces (objects $a$ and $b$ in the feet of the structured cospan): one ($a$) containing stock $I$, a sum auxiliary variable $N$, and a link from $I$ to $N$, and another ($b$) containing stock $S$, a sum auxiliary variable $N$, and a link from $S$ to $N$. The legs of the structured cospan consist of the appropriate ``inclusion" mappings (homomorphisms), which serve to send an element (e.g., variable) in the foot to the corresponding element in the apex.

\begin{figure}[!h]
\centering
\includegraphics[width=0.8\textwidth]{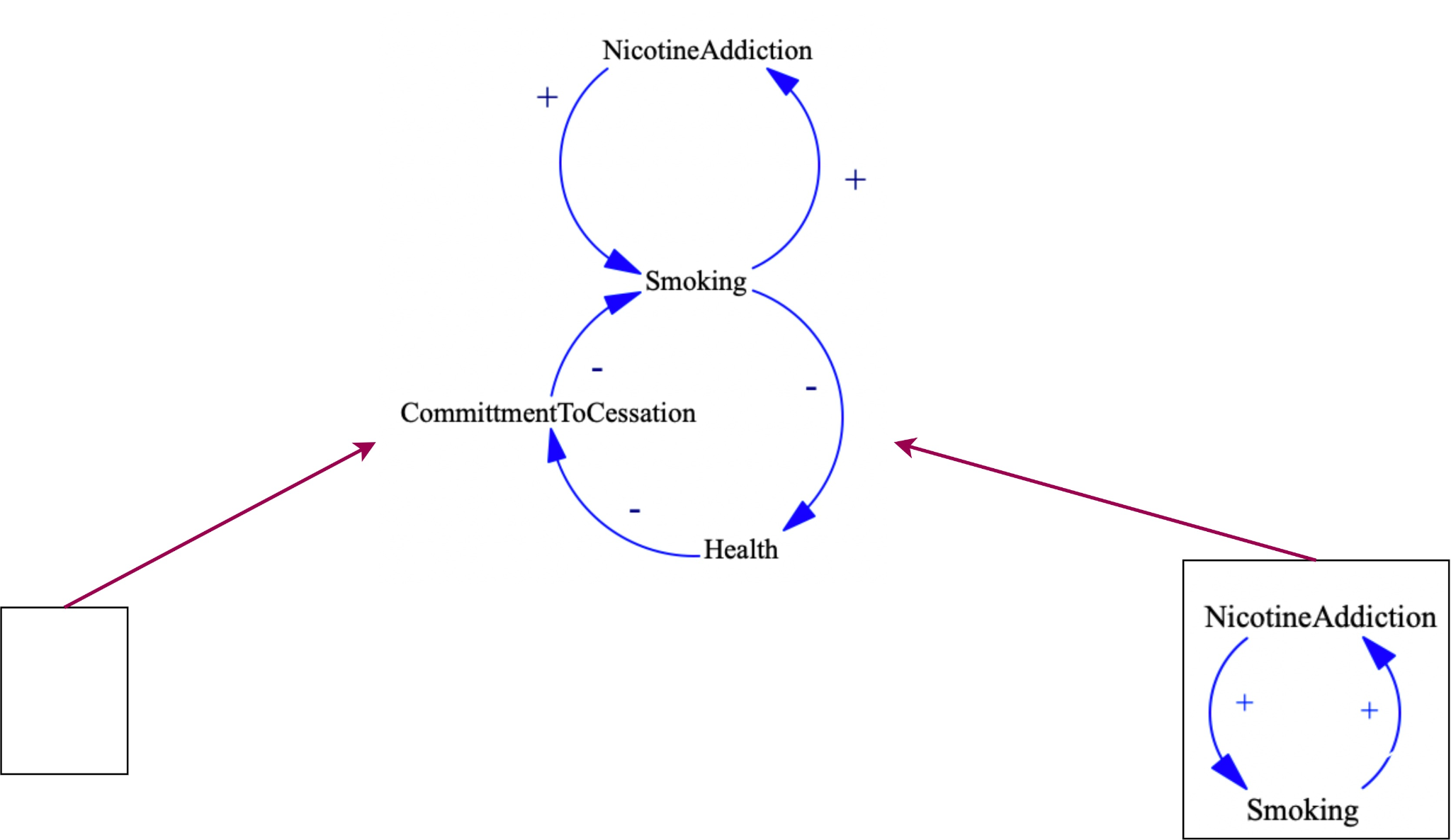}
\caption{An example of an open causal loop diagram.}
\label{figexopenclds}
\end{figure}

Figure~\ref{figexopenclds} shows an example of an open causal loop diagram providing two interfaces (bottom left and right) to a smoking causal loop diagram (the object $x$ at the apex of the structured cospan -- Figure~\ref{figstrcospan}). Within the figure, the left interface in the structured cospan (which we term $a$) is empty, while right interface (termed $b$) represents a sub-diagram consisting of variables like ``Nicotine Addiction" and ``Smoking" with positive links between them.

\subsubsection{A Composition Framework using \textit{pushout}}

\paragraph{Mathematical Basis of Composition}

\begin{figure}[!h]
\centering
\includegraphics[width=0.35\textwidth]{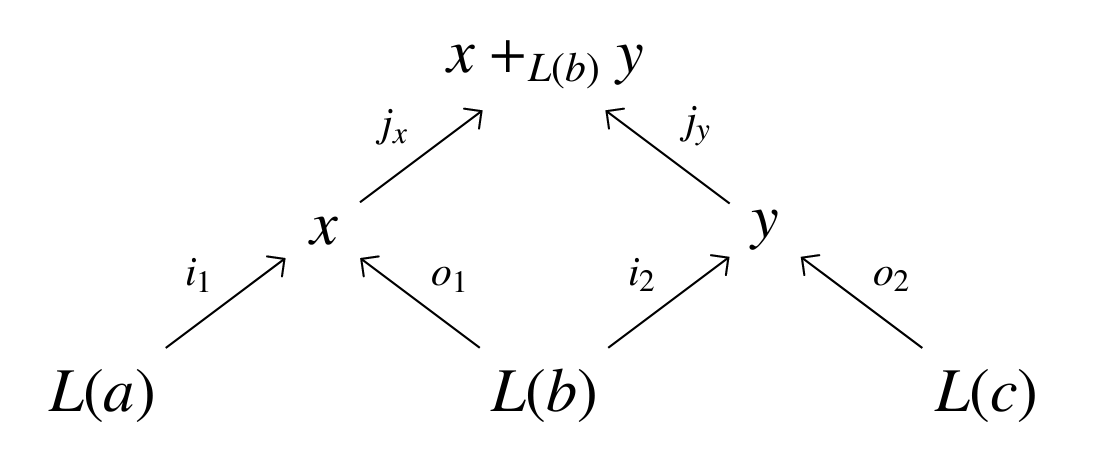}
\caption{The mathematical structure for composing diagrams $x$ and $y$ using \texttt{pushout}.}
\label{figPushout}
\end{figure}

The composition of diagrams---via composition of structured cospans representing such diagrams in a fashion that exposes interfaces to them---is formalized through a mathematical operation known as a \textit{pushout} \cite{SevenSketches}, a core concept in category theory, and one distinguished by ``universal properties" that allow it to be defined in diverse contexts. This operation enables two diagrams to be ``glued" together by aligning shared interfaces. 

Figure~\ref{figPushout} illustrates the mathematical structure of composing diagrams using the categorical concept of \texttt{pushout}. The interpretation of composition is as follows: if we have two structured cospans ($L(a)\rightarrow x \leftarrow L(b)$ and $L(b)\rightarrow y \leftarrow L(c)$), where a diagram $x$ is open with two interfaces ($a$ and $b$) and a diagram $y$ is also open with two interfaces ($b$ and $c$), and where each of such interfaces has simple ``inclusion maps" into the corresponding diagram in the apices ($x$ or $y$), we can compose diagrams $x$ and $y$ by ``gluing" them together at points identified by their common interface ($b$). Doing so means that we identify the subparts represented in the common interface ($b$) and create a composed diagram ($x_{L(b)}y$). The resulting composed diagram ($x_{L(b)}y$) is obtained by ``gluing $x$ and $y$ together" through their shared interface, and creating an overall composed diagram that ensures consistency and preserves the structural properties of the individual diagrams.  More specifically, if an element in the common interface (say, a stock) points to a given element $e_x$ of diagram $x$, and an element $e_y$ in diagram $y$, while the composed diagram will include all elements (stocks, flows, links, etc.) from diagram $x$ and all of those from diagram $y$, the stock playing the role of $e_x$ and $e_y$ will be one and the same.

\paragraph{Examples of Composing Diagrams}

\begin{figure}[!h]
\centering
\includegraphics[width=1\textwidth]{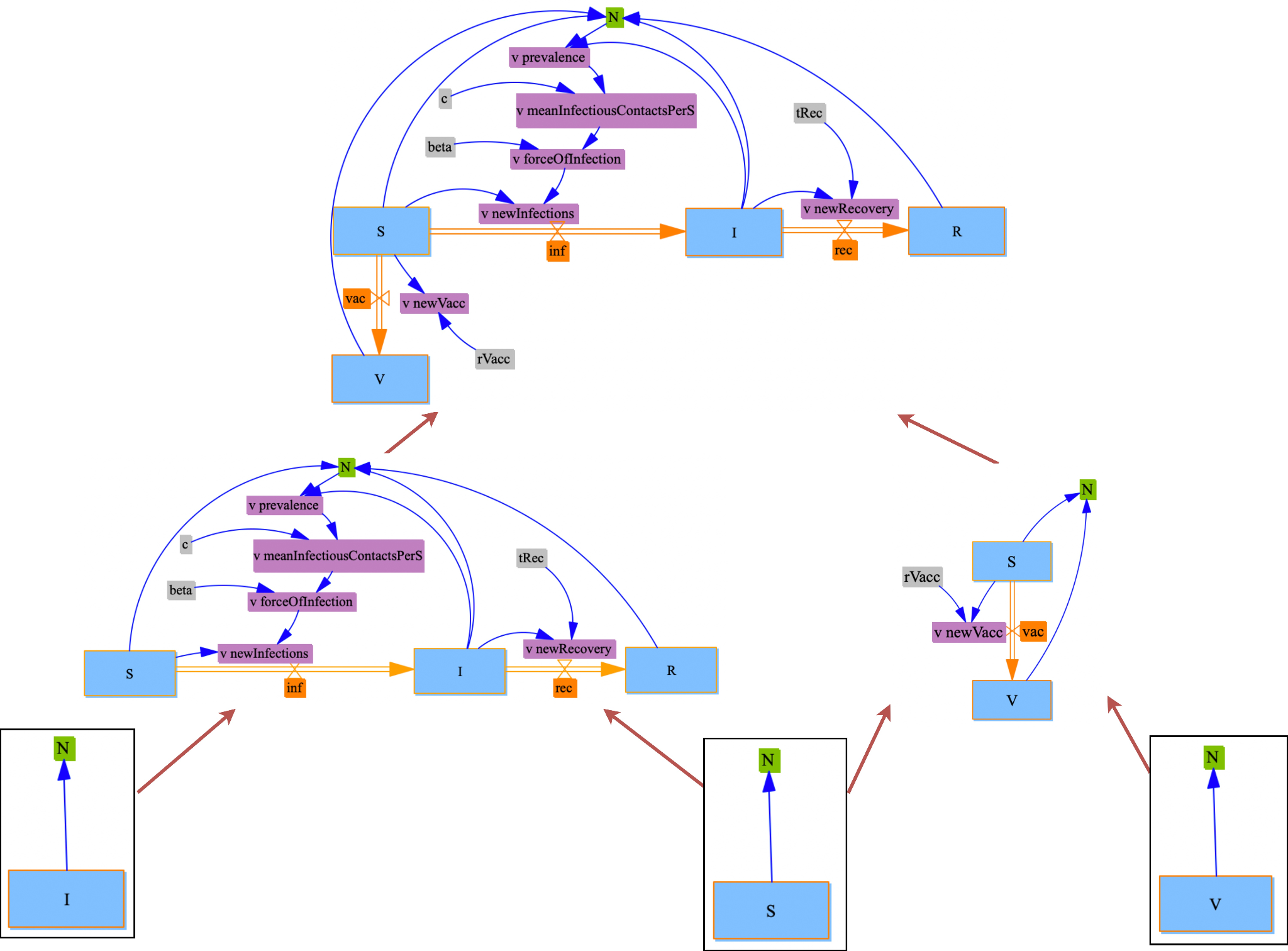}
\caption{An example of composing stock \& flow diagrams.}
\label{figexComposeSFDs}
\end{figure}

\begin{figure}[!h]
\centering
\includegraphics[width=1\textwidth]{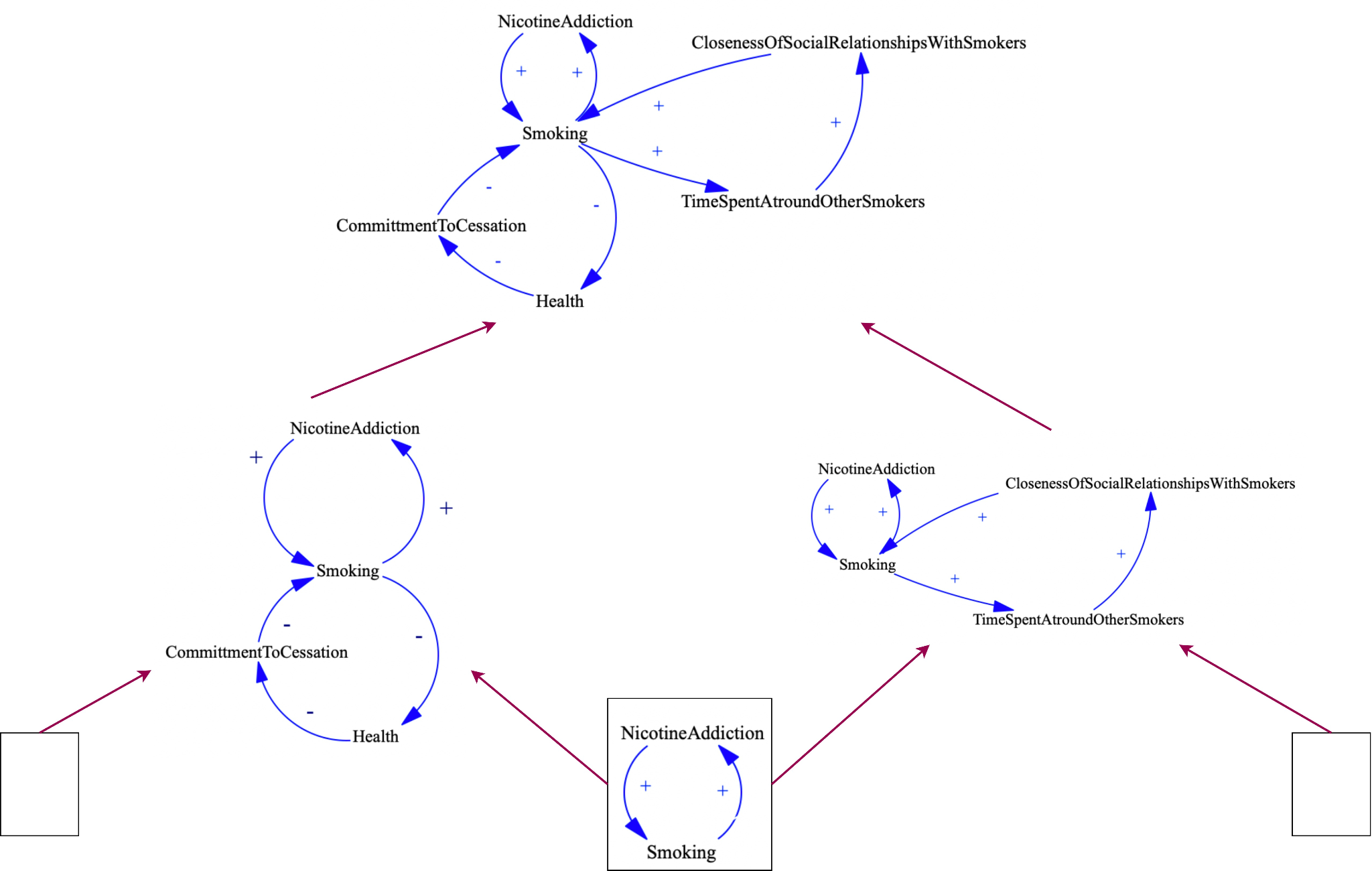}
\caption{An example of composing causal loop diagrams.}
\label{figexComposeCLDs}
\end{figure}

Figure~\ref{figexComposeSFDs} and Figure~\ref{figexComposeCLDs} show two examples of composing diagrams using the categorical operation of pushout. 

Figure~\ref{figexComposeSFDs} shows a how stock \& flow diagram representing a vaccination process (denoted $y$ in Figure~\ref{figstrcospan}) can be integrated (composed) with another SIR stock \& flow diagram (denoted  $x$ in Figure~\ref{figstrcospan}) representing infection spread by identifying,  in each of the two diagrams $x$ and $y$, the counterparts of shared stock $S$ (i.e., the stock in $x$ to which $S$ maps, and that in $y$ to which $x$ maps), the counterparts of sum dynamic variable $N$, and the counterparts of the link from $S$ to $N$. 

Similarly, Figure~\ref{figexComposeCLDs} shows another example of composing two causal loop diagrams with overlapping variables and links defined in the common interfaces to form a unified representation.

This compositional approach extends readily to other types of diagrams, including system structure diagrams. In each case, the framework provides a rigorous mathematical basis for modularity, enabling the systematic and well-defined construction of larger models from smaller, equally well-defined, components. The ability to formally encode and compose diagrams in this way --- thereby producing composite diagrams that can be used in all the ways in which bespoke diagrams can be used --- supports scalability, facilitates collaboration, and enhances the clarity and reusability of complex models.

\subsection{Applications of Homomorphisms}\label{homomorphism}

Homomorphisms play a central role in this framework, by characterizing structure-preserving mappings between diagrams. Intuitively, a homomorphism allows us to relate the components and structure of one diagram to another, while preserving their essential relationships. This concept --- which also appeared in the context of structured cospans --- is particularly useful in System Dynamics modeling, where identifying patterns and model stratification are key tasks.

While homomorphisms played an important supporting role when composing diagrams, this section further discusses two central applications of homomorphisms: identifying patterns and assigning types to diagram elements.

\subsubsection{Identifying Patterns}\label{PatternIdentification}

\begin{figure}[!h]
\centering
\includegraphics[width=1\textwidth]{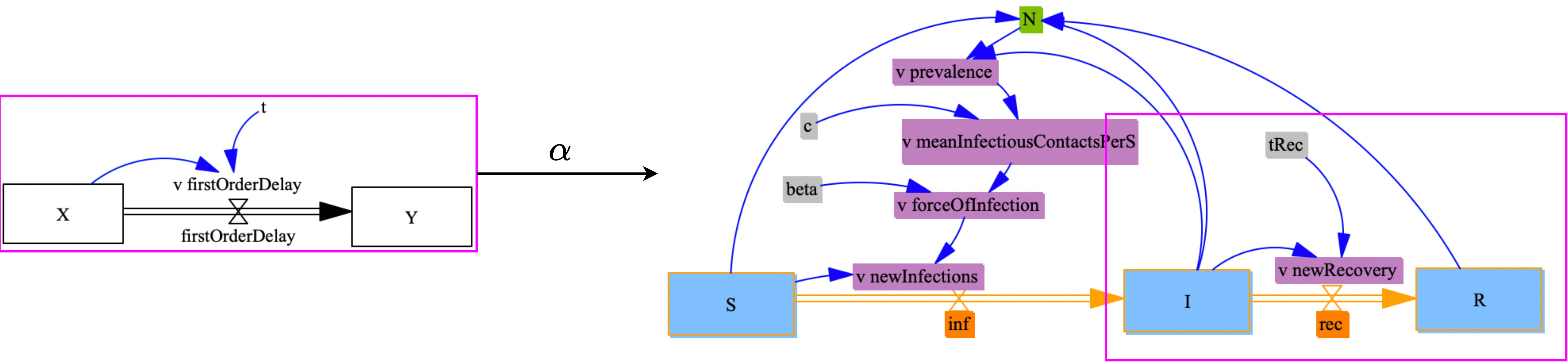}
\caption{An example of identifying a classic first-order delay pattern in an SIR stock \& flow diagram.}
\label{fighomopattern}
\end{figure}

Homomorphisms enable the identification of specific patterns within a target diagram by identifying mappings of a predefined pattern diagram into that target. For example, consider the task of identifying instances of a classic first-order delay in a stock \& flow diagram (Figure~\ref{fighomopattern}). This process involves the following steps:
\begin{enumerate}
    \item Define a pattern diagram that describes a first-order delay. For the variant considered here, this diagram includes two stocks (\(X\) and \(Y\)) connected by a flow governed by a constant parameter \(t\), representing the delay.
    \item Search for one or more homomorphisms, denoted as \(\alpha\), mapping components of the pattern diagram to corresponding components in the target diagram. For instance:
    \begin{itemize}
        \item Stock \(X\) in the pattern diagram is mapped to stock \(I\) in the target diagram.
        \item Stock \(Y\) is mapped to stock \(R\).
        \item The first-order delay flow is mapped to the recovery flow.
        \item The constant parameter \(t\) is mapped to the recovery delay parameter \(tRec\).
    \end{itemize}
    \item For each such homomorphism identified, record the substructures in the target diagram that match the pattern within the homomorphism. 
\end{enumerate}

Each homomorphism corresponds to one instance of the pattern in the target diagram.  For complex diagrams, there may be multiple instances of the pattern. In such cases, multiple homomorphisms can be identified, each identifying a specific occurrence of the first-order delay. This approach provides a systematic way to detect and analyze recurring patterns within diagrams; in Section \ref{signedCat} below, we note how recent advances employing double categories allow for greater generality and flexibility in pattern matching.

\subsubsection{Assigning Types to Diagram Elements}

\begin{figure}[!h]
\centering
\includegraphics[width=1\textwidth]{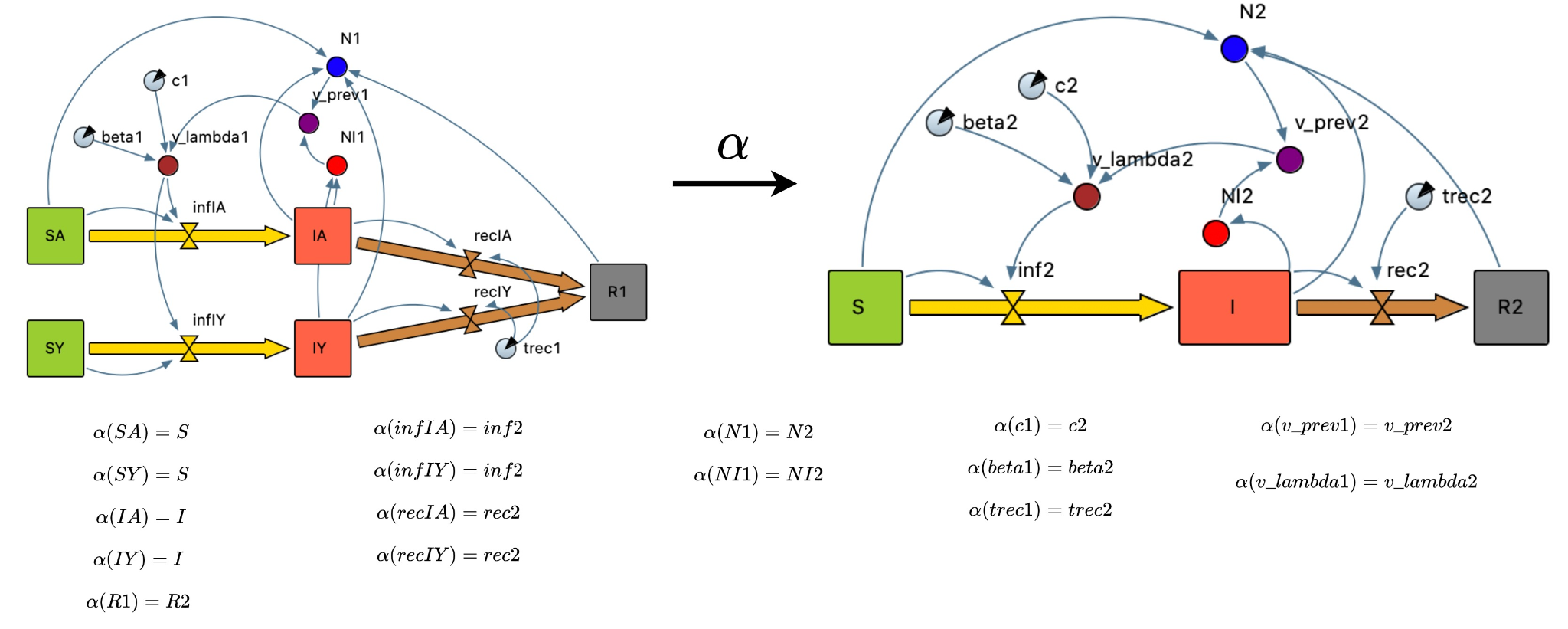}
\caption{An example of assigning types to elements in stock \& flow diagrams.}
\label{fighomo_type}
\end{figure}

In a process that plays a central role in stratification of stock \& flow diagrams discussed in the next section, homomorphisms can further be used to assign types to elements in such diagrams by mapping them to a predefined ``type diagram" (Figure~\ref{fighomo_type}). 

This process of assigning types to a source stock \& flow diagram involves the following:
\begin{enumerate}
    \item Define a type diagram where each component is associated with a specific type (the target diagrams in this example).
    \item Using a homomorphism, we map elements in the source stock \& flow diagram to the target ``type diagram" defined in the previous step. The details of the mapping of each element of the homomorphism $\alpha$ in this example are listed in Figure~\ref{fighomo_type}.
    \item Read out the type for each element in the source diagram based on its mapping to the type diagram. Visualization techniques, such as color coding, can be used to highlight elements with the same type.
\end{enumerate}

This type-assignment approach provides a formal way to categorize and analyze diagram elements, ensuring consistency across different representations.  

Homomorphisms serve very useful roles not only for stock \& flow diagrams but also for causal loop diagrams and system structure diagrams. By defining structure-preserving mappings, homomorphisms can be used to identify patterns and assign types in various diagrammatic representations, including causal loop diagrams and system structure diagrams. This flexibility highlights the power of homomorphisms in enabling modular, scalable, and systematic reasoning about complex systems using System Dynamics diagrams.

\subsection{A Stratification Framework}\label{Stratification}

The concept of ``type diagrams" serves as the foundation of the stratification framework. This stratification framework enables the creation of complex stratified models by systematically dividing aggregate diagrams into subgroups based on defined characteristics, such as regions, age groups, or other stratification structures.  It bears emphasis that such stratification by some characteristic here does not simply impose a new dimension of layering within the model to which it is applied. Rather, it can be used to distinguish not only different states that apply for that characteristic, but also different kinds of \textit{logical processes} (associated with flows) that lead to changes among such states.  For example, in a lifecourse model of weight change and cardiovascular disease (CVD), we can distinguish weight-related states and the flows between them from those involved in the natural history of cardiovascular disease. In a similar manner, we could further delineate processes that change weight- and CVD-related states from those associated with age categories and the processes of aging.  Each of those three kinds of concerns --- weight change, cardiovascular disease, and aging --- could be associated with a distinct dimension of stratification.

Both stratification and the capacity to compose a model out of subpieces support achieving a separation of concerns in model construction. Stratification differs from composition by focusing on subdividing an aggregate model into finer --- often fairly independent ---submodels while maintaining consistency through mathematical rigor.   The relatively independent nature of the processes depicted across different dimensions of stratification allow it to be characterized well by another universal construction in category theory --- the \textit{pullback}.

\subsubsection{Mathematical Basis for Stratification}

\begin{figure}[!h]
\centering
\includegraphics[width=0.4\textwidth]{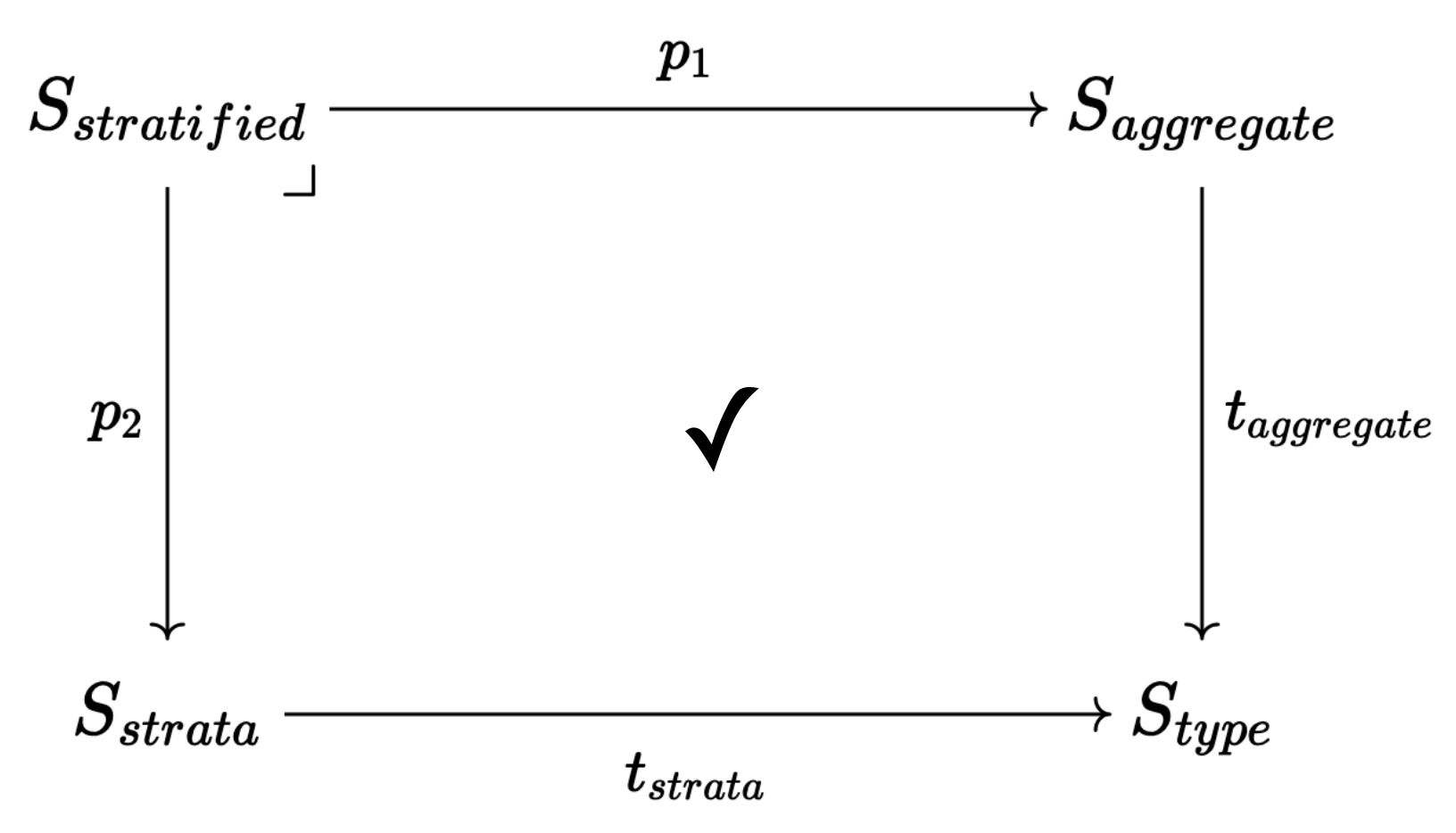}
\caption{The mathematical structure of a \texttt{pullback} square.}
\label{figpullback}
\end{figure}

The stratification framework leverages the concept of \texttt{pullback} in category theory \cite{SevenSketches}. While the pullback is another categorical construction distinguished by universal properties that allow it to be applied in a vast array of categories, the pullback of concern here operates within a category of diagrams -- the category of (closed) diagrams introduced in Section \ref{Subsection:closedDiagrams} for each of causal loop diagrams, system structure diagrams, and stock \& flow diagrams. 

Figure~\ref{figpullback} shows the ``pullback square" that forms the mathematical context for characterizing stratified diagrams using a pullback operation. The interpretation of the pullback square is that \(S_{\text{aggregate}}\), \(S_{\text{strata}}\), \(S_{\text{type}}\), and \(S_{\text{stratified}}\) are objects (instances of diagrams) in a category of (closed) diagrams defined introduced in Section \ref{Subsection:closedDiagrams}. The interpretation of each object is given as follows:
\begin{itemize}
    \item \(S_{\text{aggregate}}\): A diagram representing the system absent stratification (e.g., for the entire population).
    \item \(S_{\text{strata}}\): A diagram capturing states and processes along the desired stratification dimension (e.g., age-based stratification into the three categories of children, adults, and seniors, with corresponding aging flows).
    \item \(S_{\text{type}}\): A type diagram where each component represents a specific type.
    \item \(S_{\text{stratified}}\): The stratified model generated by combining the diagrams of \(S_{\text{aggregate}}\) and \(S_{\text{strata}}\) using the pullback operation on the square depicted in Figure~\ref{figpullback}.
\end{itemize}

The pullback square is bounded by 4 diagram homomorphisms: $p_1$, $p_2$ (which are given as part of the formalism as projections), and context-specific homomorphisms (\(t_{\text{aggregate}}\) and \(t_{\text{strata}}\)).  The latter two homomorphisms assign types to the components in \(S_{\text{aggregate}}\) and \(S_{\text{strata}}\), respectively, by mapping each to the type diagram \(S_{\text{type}}\).  The definitions of these homomorphisms helps ensure consistency and can enables automatic generation of the stratified model.  It bears emphasis that while the resulting (pullback) model \(S_{\text{stratified}}\) will exhibit combinations of structures drawn from \(S_{\text{aggregate}}\) and \(S_{\text{strata}}\), the fact that this operation is specified by two separate separate morphisms allows this combinatorial structure to be described in a modular fashion.  The advantages of this modular approach builds as additional dimensions dimensions of stratification are considered in the pullback.

\begin{figure}[!h]
\centering
\includegraphics[width=0.38\textwidth]{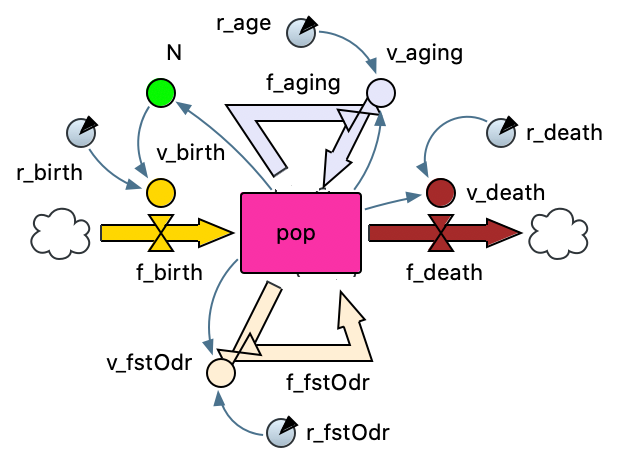}
\caption{An example of a type stock \& flow diagram (\(S_{\text{type}}\)).}
\label{figtypediagram}
\end{figure}

\begin{figure}[!h]
\centering
\includegraphics[width=0.5\textwidth]{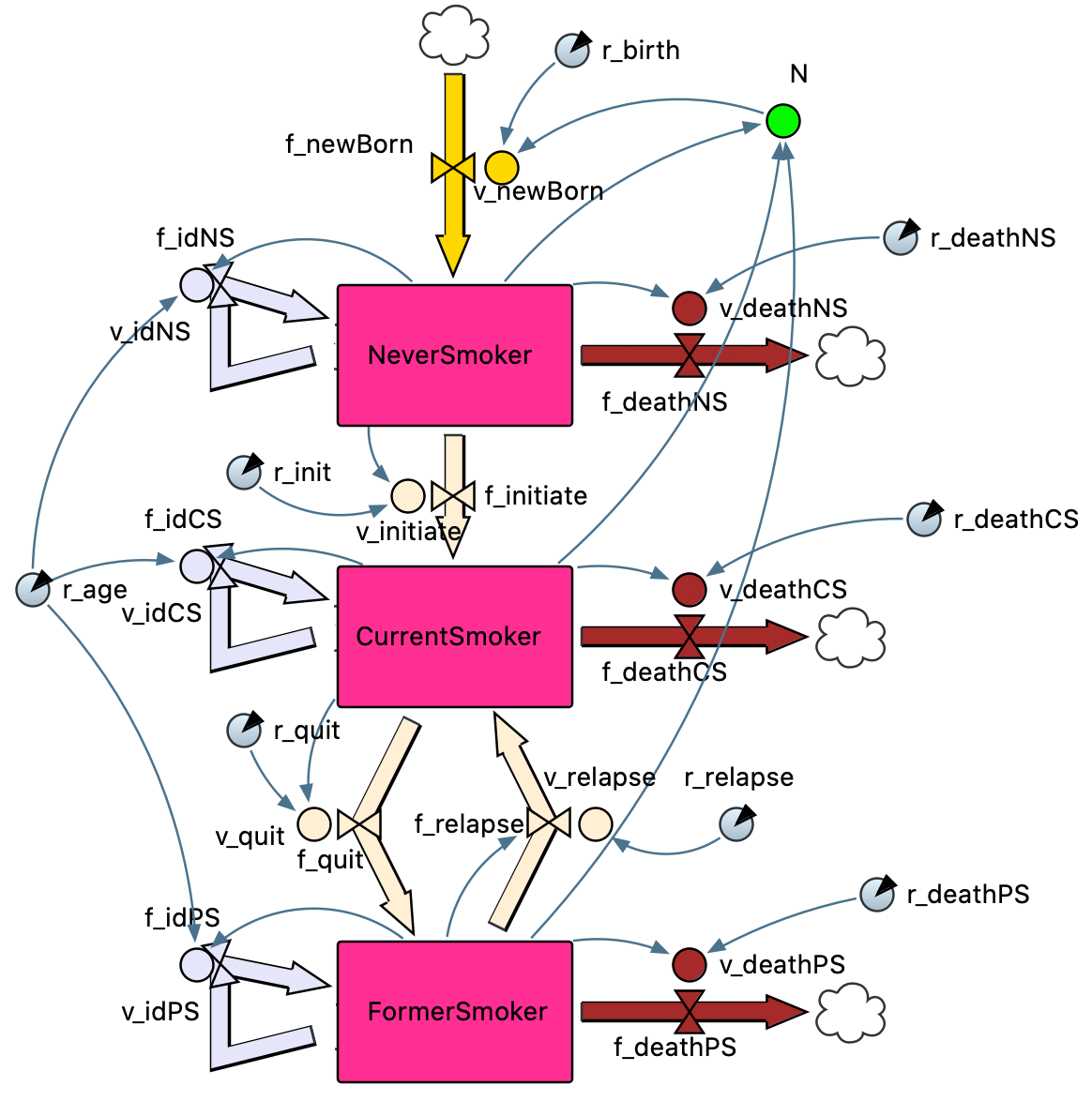}
\caption{An example of aggregate population smoking stock \& flow diagram (\(S_{\text{aggregate}}\)).}
\label{figSaggregate}
\end{figure}

\begin{figure}[!h]
\centering
\includegraphics[width=0.6\textwidth]{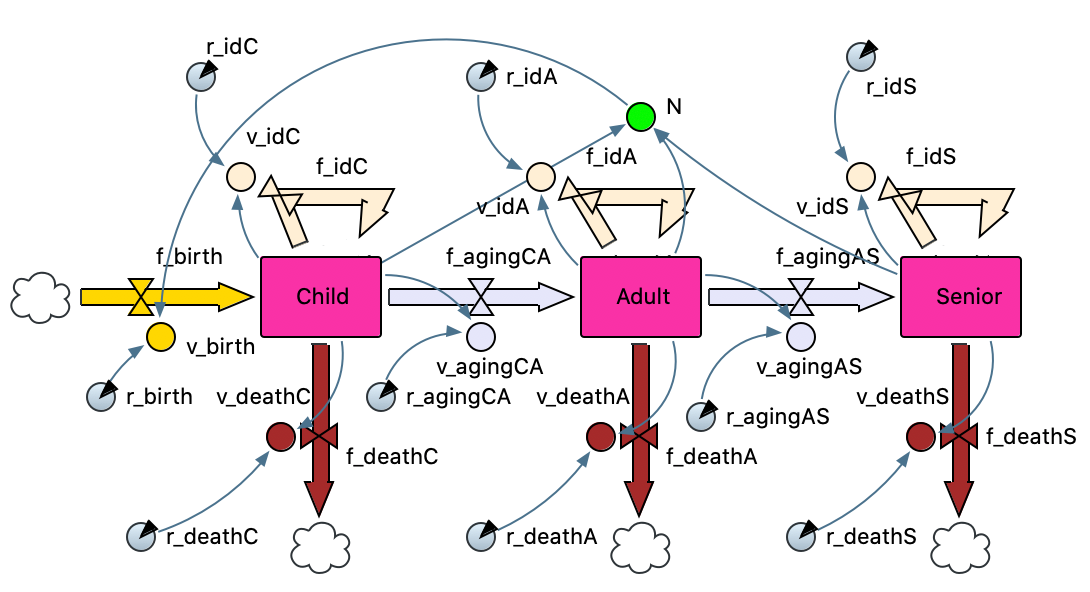}
\caption{An example of stock \& flow diagram (\(S_{\text{strata}}\)) representing a stratification dimension -- here, capturing age subgroup structure.}
\label{figSStrata}
\end{figure}

\begin{figure}[!h]
\centering
\includegraphics[width=1\textwidth]{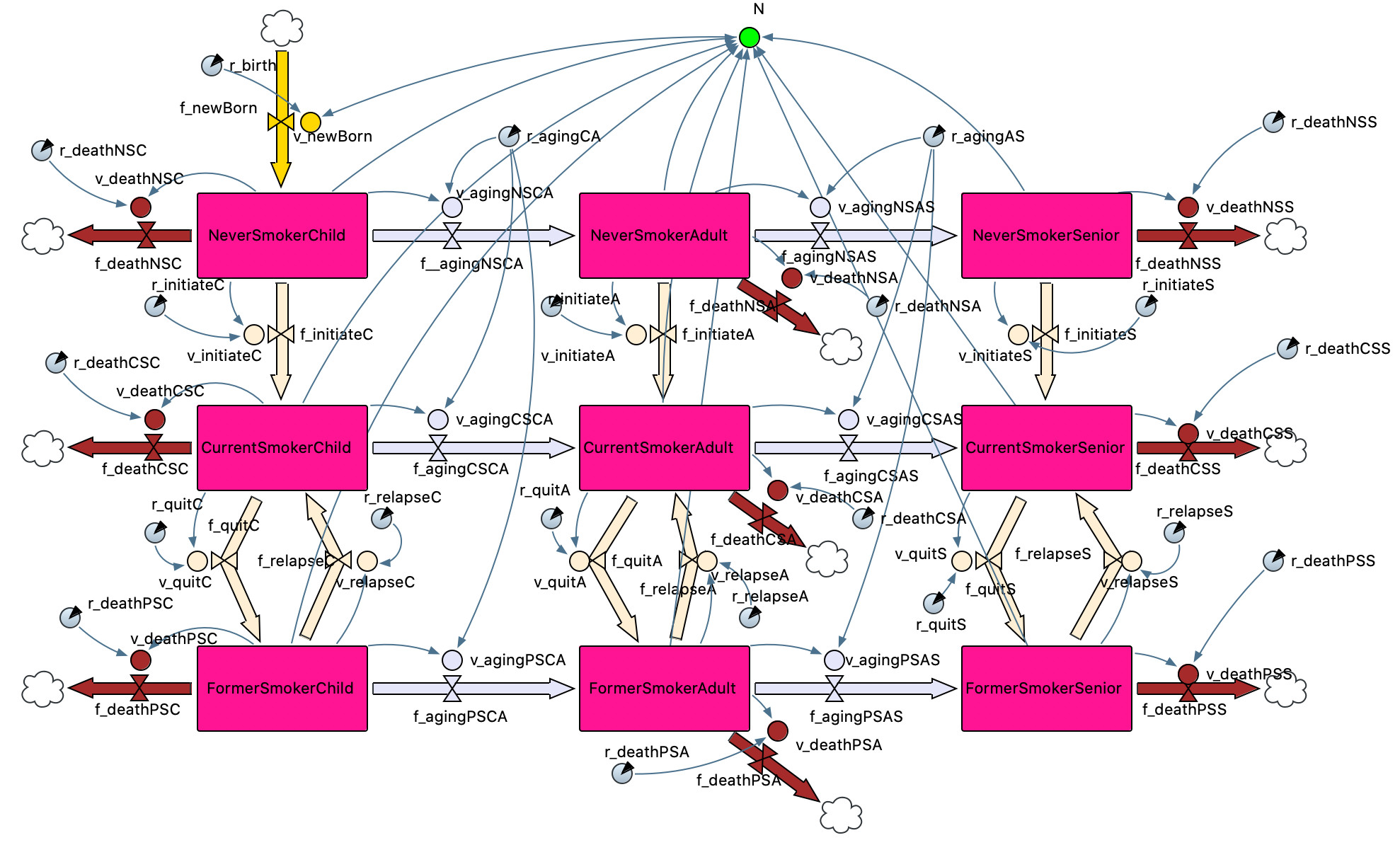}
\caption{An example of smoking stock \& flow diagram stratified using \texttt{pullback} (\(S_{\text{stratified}}\)).}
\label{figSStratified}
\end{figure}

\subsubsection{Example: Smoking Model}

Figure~\ref{figSStratified} shows an example of a smoking stock \& flow diagram stratified into three subgroups: \textit{child}, \textit{adult}, and \textit{senior}. This stratified diagram can be obtained by using the pullback square (Figure~\ref{figpullback}), which involves taking the product of the aggregate population diagram (Figure~\ref{figSaggregate}) representing the tobacco-use behavior structure and the diagram representing the structure among different age subgroups (Figure~\ref{figSStrata}), based on the type diagram (Figure~\ref{figtypediagram}).  It is to be emphasized that such stratification supports incorporating both the states (captured as stocks) and the processes that change such states (expressed as flows) in each stratum.

The main idea of constructing stratified diagrams is as follows: each component in the aggregate diagram (Figure~\ref{figSaggregate}) is stratified into the number of elements in the strata diagram (Figure~\ref{figSStrata}), provided they share the same type, as determined by each of their mappings to the type diagram (Figure~\ref{figtypediagram}). For visualization purposes, this paper uses the colors to represent the types defined in the type diagram (Figure~\ref{figtypediagram}); thus, for an item (e.g., flow) in the aggregate diagram of a given color, stratification leads that item to be applied for each item of the same color within the strata diagram (thus, a corresponding flow is provided in the strata diagram).

This stratification framework provides a systematic way to: automatically generate complex stratified models, preserve consistency between aggregate and stratified models, and enable scalability by modularly combining small, well-defined diagrams. By leveraging category theory and pullback squares, this stratification framework ensures mathematical rigor and provides a powerful tool for creating stratified System Dynamics models.

\subsection{Structure Preserving Mappings Between Different Types of Diagrams}\label{Map}

We have seen the structure-preserving mappings between diagrams of the same type termed \textit{homomorphisms} in Section~\ref{homomorphism}. In this section, we discuss structure-preserving mappings among different types of System Dynamics diagrams --- such as from stock \& flow diagrams to causal loop diagrams, or system structure diagrams to causal loop diagrams. Such mappings provide a rigorous mathematical approach for abstracting diagrams, finding analogues for instances of more textured types of diagrams within instances of more abstract diagrams, as well as transforming and analyzing relationships across different diagram types. This approach enhances our ability to model complex systems using System Dynamics by combining insights from various perspectives, and by keeping diagrams articulated at higher levels of abstraction in sync with the evolution of more granular diagrams.

\begin{figure}[!h]
\centering
\includegraphics[width=0.5\textwidth]{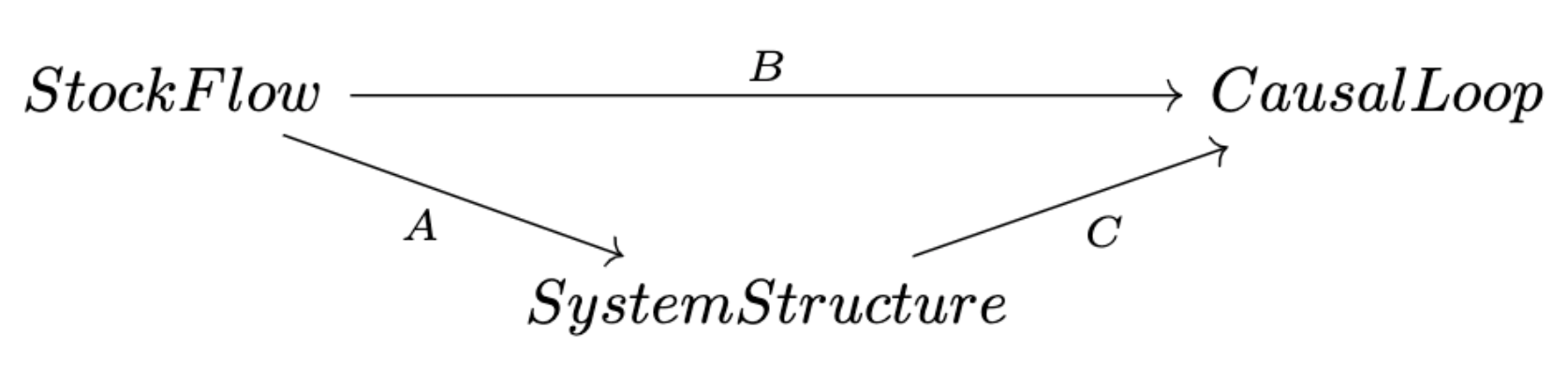}
\caption{Types of structure-preserving mappings among the three types of diagrams.}
\label{figFunctorMappings}
\end{figure}

Figure~\ref{figFunctorMappings} characterizes types of structure-preserving mappings among the three types of diagrams. Mathematically, arrows $A$, $B$ and $C$ in Figure~\ref{figFunctorMappings} represent \textit{functors} among those three categories of (closed) diagrams (introduced in Section \ref{Subsection:closedDiagrams}).  We now turn to discuss each of these functors. 

\subsubsection{Mapping System Structure Diagrams to Causal Loop Diagrams}

\begin{figure}[!h]
\centering
\includegraphics[width=0.8\textwidth]{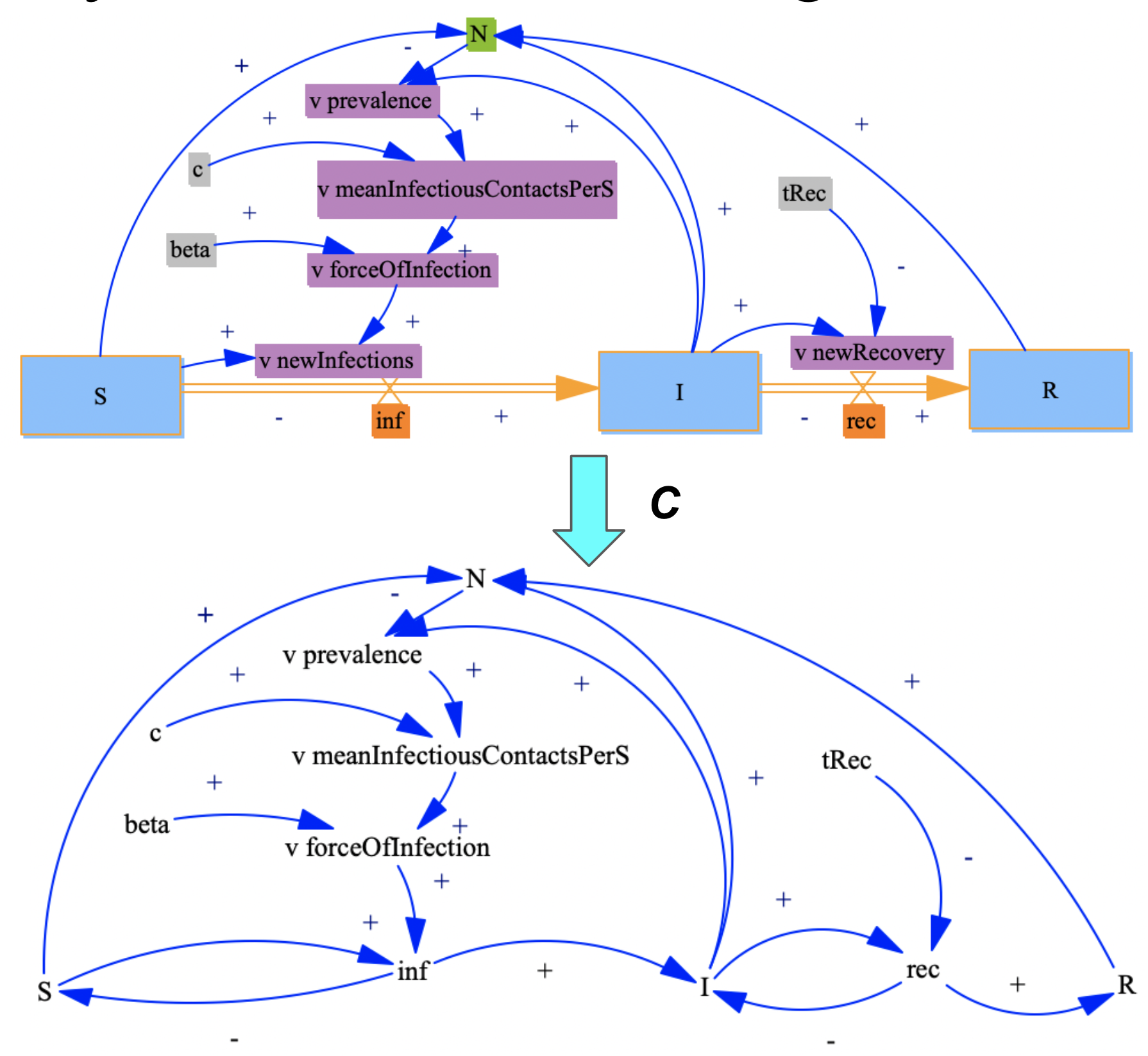}
\caption{An example of functor mapping from a SIR system structure diagram to a corresponding causal loop diagram.}
\label{figexSSD_CLD}
\end{figure}

The mapping from system structure diagrams to causal loop diagrams ($C$) can be defined using functorial data migration \cite{aduddell2024compositional, spivak2025functorial, SPIVAK201231} principles from applied category theory. The essential idea is that the functor mapping from the category of system structure diagrams (SSDs) to the category of causal loop diagrams (CLDs), denoted as $C: SystemStructure \rightarrow CausalLoop$, can be automatically generated from the functor mapping from the schema category of causal loop diagrams ($SchCLDs$) to the category of disjoint union of database queries on the schema category of system structure diagrams ($SchSSDs$):

\begin{align*}
V &\rightarrow S + SV + V + P \\
L &\rightarrow LV + LS + LSV + LPV + LVV + I + O \\
A &\rightarrow A \\
\text{src} &\rightarrow [\text{lvs}, \text{lss}, \text{lsvsv}, \text{lpvp}, \text{lvsrc}, \text{fv}\circ\text{ifn}, \text{fv}\circ\text{ofn}] \\
\text{tgt} &\rightarrow [\text{lvv}, \text{lssv}, \text{lsvv}, \text{lpvv}, \text{lvtgt}, \text{is}, \text{os}] \\
\text{polarity} &\rightarrow [\text{polarityLV}, \text{positiveLS}, \text{polarityLSV}, \text{polarityLPV},\\ &\text{polarityLVV}, \text{positiveI}, \text{negativeO}]
\end{align*}

Once these mappings are defined at the schema level, they can automatically generate mappings ($C$) from specific system structure diagrams to corresponding causal loop diagrams. Figure~\ref{figexSSD_CLD} shows an example of using functor mapping $C$ to convert an SIR system structure diagram to a corresponding SIR causal loop diagram automatically. Using the SIR model shown in Figure~\ref{figexSSD_CLD} as an example, the interpretation of the functor mapping of $C$ (refer to the above box) is as follows. If we view the instance of both the SIR system structure diagram and the SIR causal loop diagram as encoded categorical databases:

\begin{itemize}
    \item The structured table of dynamic variables \texttt{V} in the SIR causal loop diagram is the \textit{Union} of the structured tables of stocks \texttt{S}, sum auxiliary variables \texttt{SV}, auxiliary variables \texttt{V}, constant parameters \texttt{P}, and flows \texttt{F} in the SIR system structure diagram, with duplicate rows removed. Notably, since the identity morphism for object $V$ in $SchCLDs$ ($id_V$) maps to the morphism $fv$ in $SchSSDs$, it indicates that each flow and its corresponding auxiliary variable are identified as one dynamic variable in causal loop diagrams.
    \item The attributes table of \texttt{A} in the causal loop diagram is the same as the attributes table of \texttt{A} in the system structure diagram.
    \item The structured table of links \texttt{L} in the SIR causal loop diagram is the \textit{Union} of the structured tables of different types of links (\texttt{LV}, \texttt{LS}, \texttt{LSV}, \texttt{LPV}, \texttt{LVV}), the inflows \texttt{I}, and outflows \texttt{O} of the system structure diagram. The three columns ``src," ``tgt," and ``polarity" are also mapped by the corresponding columns in those tables in the system structure diagram (details visible in the above box).
\end{itemize}

\subsubsection{Mapping Stock \& Flow Diagrams to System Structure Diagrams}

\begin{figure}[!h]
\centering
\includegraphics[width=0.8\textwidth]{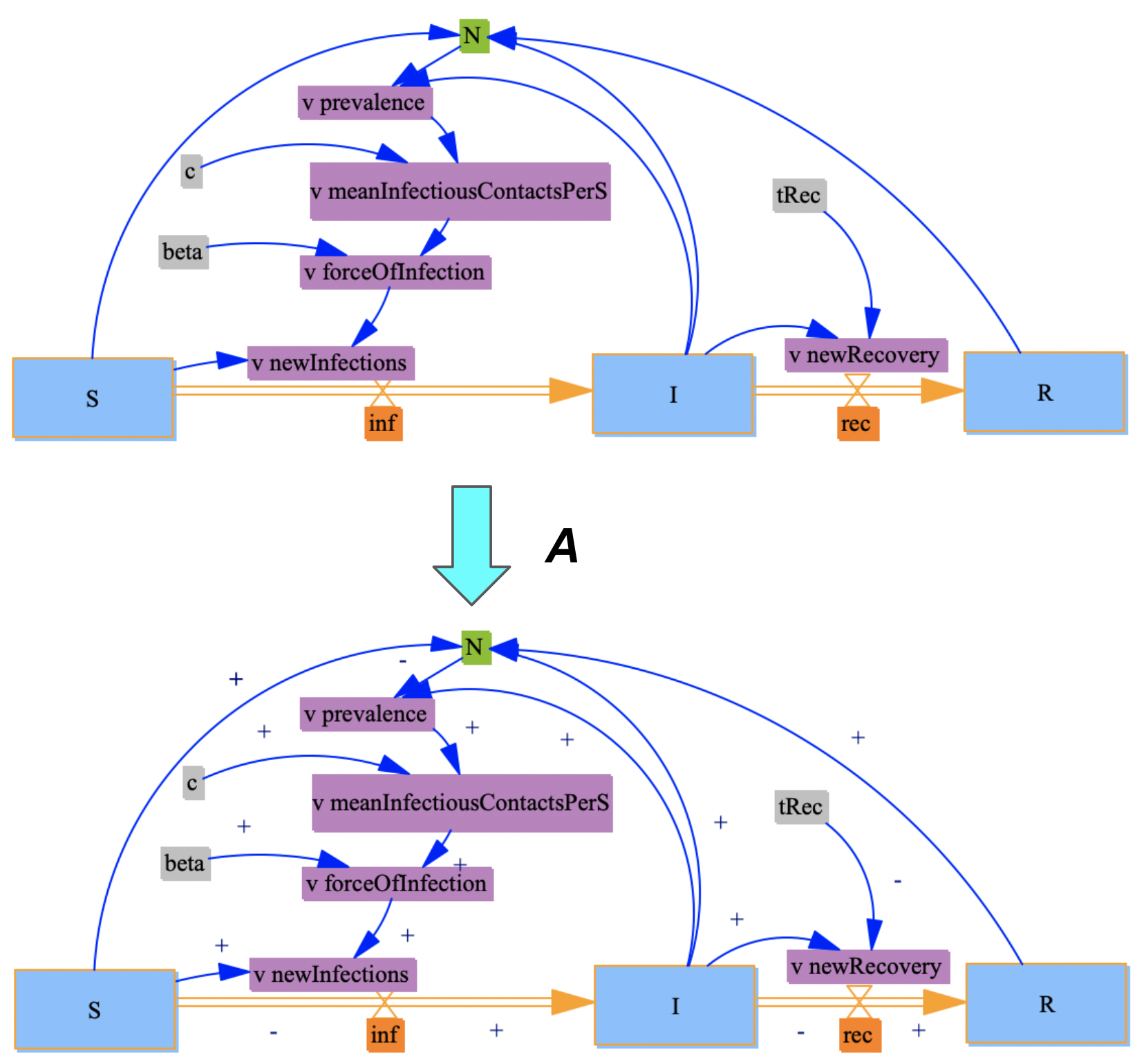}
\caption{An example of mapping from an SIR stock and flow diagram to a corresponding system structure diagram.}
\label{figexSFD_SSD}
\end{figure}

The operation of a structure-preserving mapping from stock \& flow diagrams to system structure diagrams ($A$) is intuitive due to the structural similarity between the two types of diagrams. Figure~\ref{figexSFD_SSD} illustrates an example of using the mapping $A$, which maps an SIR stock \& flow diagram to the corresponding SIR system structure diagram:
\begin{itemize}
    \item Stocks, flows, dynamic variables, and links in stock \& flow diagrams correspond directly to their counterparts in system structure diagrams.
    \item The primary distinction lies in the attributes: system structure diagrams include polarities for links. These polarities can be derived from a combination of the structural features of the model (per reasoning in \ref{SSDPolarityReasoning}) or from the formulas used to define auxiliary (dynamic) variables in stock \& flow diagrams.
    \begin{itemize}
        \item Links (\texttt{SV}) from stocks to auxiliary variables representing sums are always assigned a positive polarity.
        \item Links from flows to stocks are assigned positive polarity for inflows \texttt{I} and negative polarity for outflows \texttt{O}.
        \item Other link polarities are derived based on the mathematical formulas of the auxiliary variables. For example, if a formula involves division (e.g., \(v_{prevalence} = I / N\)), the polarity of the link from \(I\) is positive, while the polarity of the link from \(N\) is negative, given an assumption that both the arguments ( $I$ and $N$ ) are positive values.
    \end{itemize}
\end{itemize}

This systematic process ensures that polarities in system structure diagrams are accurately generated from stock \& flow diagrams in a rigorous mathematical way.

\subsubsection{Mapping Stock\& Flow Diagrams to Causal Loop Diagrams}

\begin{figure}[!h]
\centering
\includegraphics[width=0.8\textwidth]{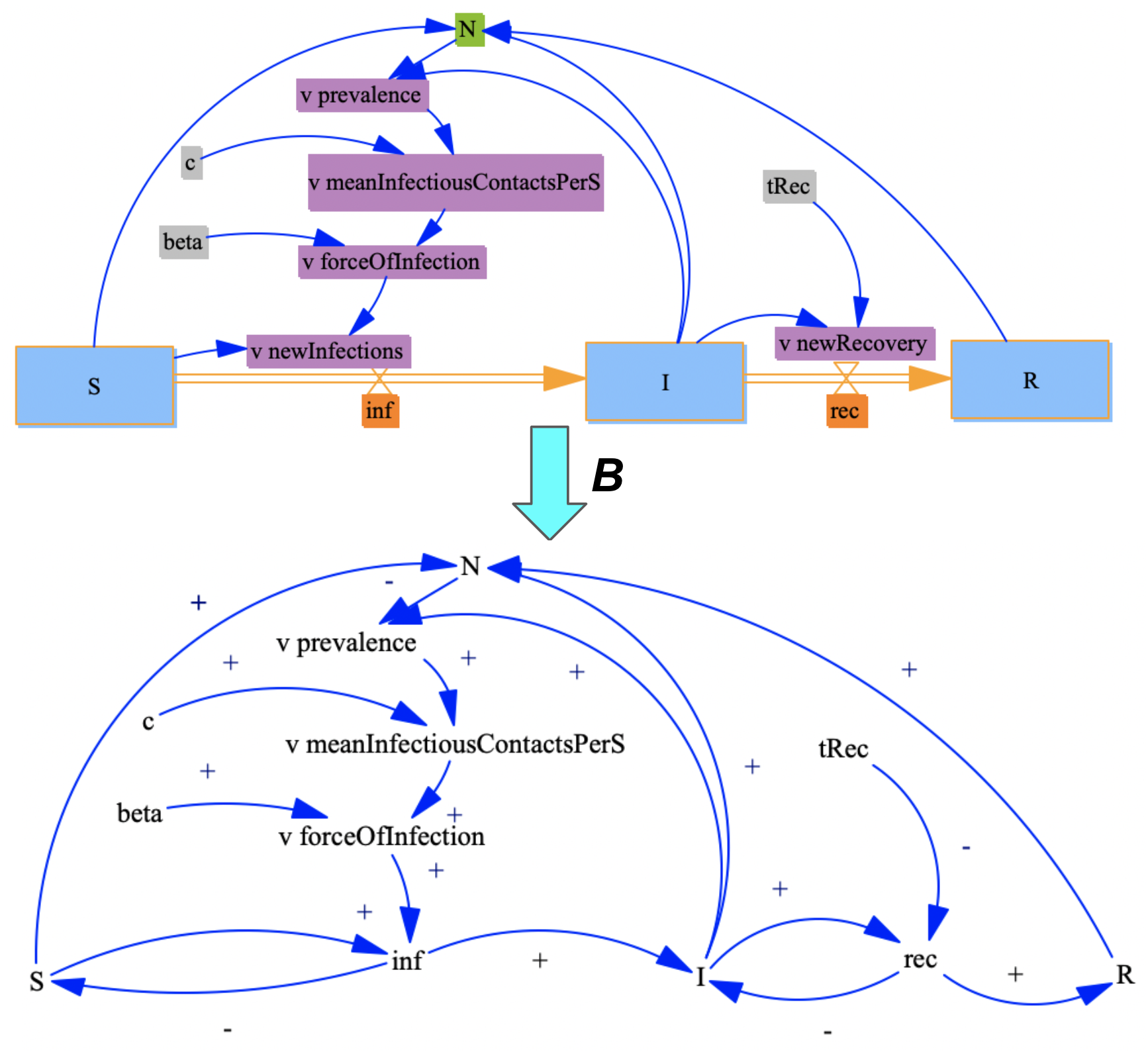}
\caption{An example of mapping from an SIR stock \& flow diagram to a causal loop diagram.}
\label{figexSFD_CLD}
\end{figure}

Because the map from a stock \& flow diagram to a system structure diagram ($A$) and the map from a system structure diagram to a causal loop diagram ($C$) each constitute a functor, they can be composed to define a map from stock \& flow diagrams to causal loop diagrams ($B$).

By composing these two mappings, any stock \& flow diagram can be transformed into a causal loop diagram, enabling analysis and integration across modeling approaches. Figure~\ref{figexSFD_CLD} shows an example of mapping from the SIR stock \& flow diagram to a corresponding causal loop diagram. Using this composition functor, we can generate the SIR causal loop diagram simply by converting the SIR stock \& flow diagram to the corresponding system structure diagram (Figure~\ref{figexSFD_SSD}) and then mapping the system structure diagram to the corresponding causal loop diagram (Figure~\ref{figexSSD_CLD}).

It further bears note that while the causal loop diagrams that result from such mappings --- such as that shown in Figure~\ref{figexSSD_CLD} --- are often finer-grained than those built by hand, this can be readily remedied by this approach.  If a coarser-grained causal loop diagram is desired, this is readily realized by following the mapping with a more coarse-grained functor mapping from a causal loop diagram to another causal loop diagram, by exploiting a categorical construct known as the category of elements.

The cross-diagram-type mappings described here allow for systematic transformations between stock \& flow diagrams, system structure diagrams, and causal loop diagrams. These mappings:
\begin{itemize}
    \item Preserve the structural integrity of diagrams,
    \item Enable integration of insights from different modeling approaches, and
    \item Facilitate automatic generation of new diagram types from existing ones.
\end{itemize}

By leveraging applied category theory and principles of data migration, this mathematical framework for mapping between diagram types provides a rigorous foundation for bridging diverse System Dynamics representations.

\subsection{Mapping Diagrams to Semantics}\label{Semantics}

\subsubsection{A Stock \& Flow Diagram Syntax and Ordinary Differential Equations Semantics Separation}

A crucial aspect of this framework is the mapping from the syntactic representation of diagrams to their corresponding semantics. As discussed earlier, stock \& flow diagrams can be viewed as the \textbf{syntactic} form of a dynamic model, representing its structure and interconnections. The \textbf{semantics}, in contrast, captures the underlying dynamics of the model,  mathematically expressed most commonly through ordinary differential equations (ODEs), but sometimes via other frameworks for characterizing dynamic behavior of state-space models.

\begin{figure}[!h]
\centering
\includegraphics[width=1\textwidth]{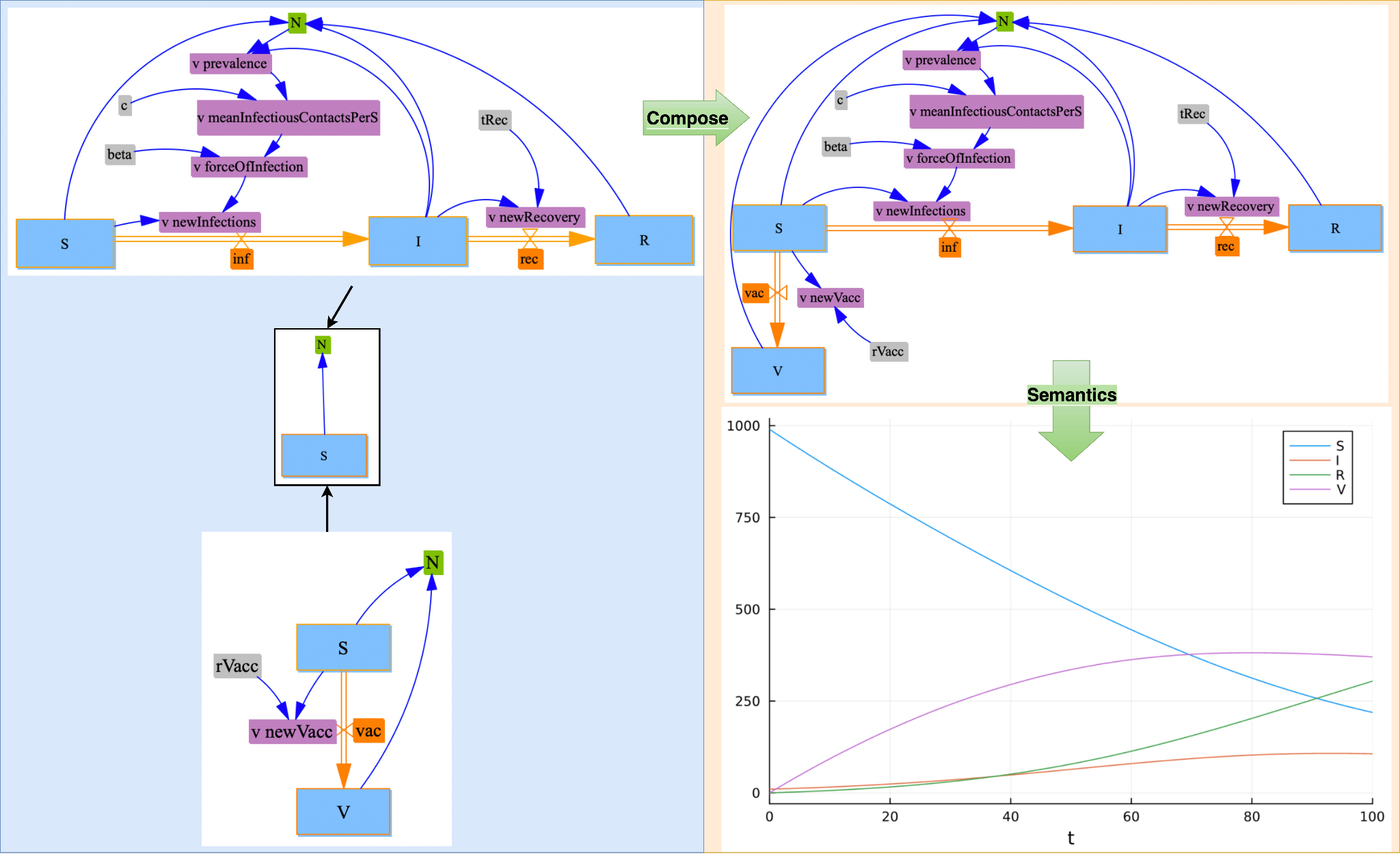}
\caption{An example of syntax versus semantics separation in a composed SIRV stock \& flow dynamic model.}
\label{figex_syntax_semantics}
\end{figure}

Figure~\ref{figex_syntax_semantics} shows an example of syntax versus semantics separation in a composed SIRV stock \& flow dynamic model. The left side of the figure illustrates the composition of two stock \& flow diagrams: the SIR stock \& flow diagram and the SV stock \& flow diagram, composed by identifying the common interface of stock \texttt{S}, sum auxiliary variable \texttt{N}, and the link from \texttt{S} to \texttt{N}. The composed SIRV stock \& flow diagram is displayed in the upper-right portion of the figure. This SIRV stock \& flow diagram represents the syntax of the corresponding SIRV dynamic model.  While this particular diagram was composed out of several pieces, the mathematically well-defined nature of stock \& flow diagram composition sketched in section \ref{Compose} ensures that the properties of the composite diagram are completely defined mathematically.  As such, the composite diagram can be readily mapped to alternative semantic domains.

The semantics of this SIRV stock \& flow diagram can be readily interpreted in a traditional fashion as ordinary differential equations (ODEs). The bottom-right part of the figure shows the line plot of the solutions to the ODEs of the dynamic system. Using the principles of System Dynamics, the mapping from the syntax of stock \& flow diagrams to the semantics of ODEs is defined as follows: for each stock in the diagram, the ODE is constructed as \[ \frac{d(\text{Stock})}{dt} = \text{Sum of Inflows} - \text{Sum of Outflows}. \] This rule is applied systematically to generate ODEs for all stocks in the diagram, ensuring a complete mathematical characterization of the model’s dynamics.

The initial values of stocks and values of constant parameters are also managed within this framework. However, they are not represented in the schema category; instead, they are assigned separately, captured by different scenarios of a Stock \& Flow model.

\subsubsection{Beyond ODEs: Alternative Semantics}
\label{sec:OtherSemantics}

While the semantics of ODEs is a traditional and well-established interpretation for stock \& flow diagrams, the categorical framework described here is designed to readily support a broader range of semantic interpretations, and the interpretation of a stock \& flow diagram as ODEs is not a privileged one. Examples already implemented within this framework include the following:
\begin{itemize}
    \item \textbf{Eigenvalue Elasticity Analysis:} In the context of Stock \& Flow diagrams, Eigenvalue Elasticity Analysis can be computed and utilized to assess the proportional sensitivity of eigenvalues of the system’s Jacobian matrix to small proportional variations in model parameters. This analysis can offer insight into the system’s stability, the emergence of oscillatory dynamics, and equilibrium behavior.
    \item \textbf{Loop Gain Analysis:} In the context of Stock \& Flow diagrams, Loop Gain Analysis can be computed and applied to evaluate the shifting dominance of different feedback loops on system evolution. By quantifying loop gains, this analysis enables the prediction, control, and optimization of dynamic systems to enhance stability and operational efficiency.
    \item \textbf{Stochastic Discrete Transitions:} Within this interpretation of a stock \& flow diagram, stocks are treated as containing discrete counts (counts of ``tokens''), and transition of a given token occurs in a memoryless stochastic process according to hazard rates (or temporal probability densities).
\end{itemize}

These alternative interpretations allow for the same syntactic form to be adapted to different contexts, enhancing the versatility of the modeling framework. This approach not only strengthens the mathematical rigor of System Dynamics modeling but also broadens its ready applicability to diverse fields and dynamic systems.  Moreover, the clean separation of syntax from semantics allows for modular mechanisms to be offered for addition of semantics of a given diagram type.


\subsection{Categorical Frontiers:  Double-Categorical Structure of Causal Loop Diagrams}\label{signedCat}

\subsubsection{Representing CLDs Using a Signed Category}
Causal loop diagrams (CLDs) play a vital role in System Dynamics by representing feedback mechanisms and their associated polarities. To enhance the analysis and utility of CLDs, we sketch here an advanced categorical structure introduced in \cite{aduddell2024compositional} that builds on fundamental principles while introducing new mathematical constructs. 

\begin{figure}[!h]
\centering
\includegraphics[width=0.4\textwidth]{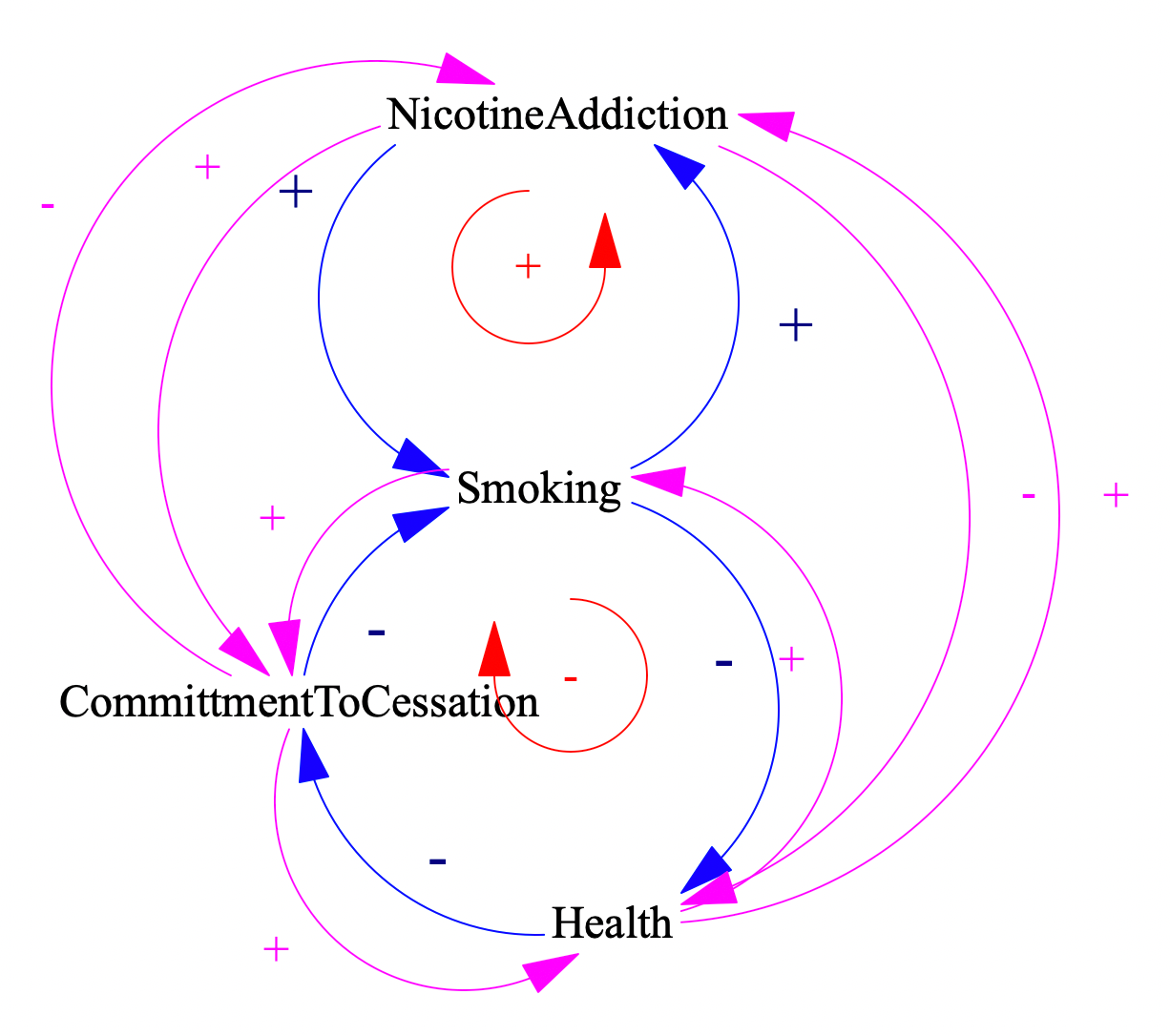}
\caption{An example of the smoking CLD supporting composition using a double category structure.}
\label{figexsignedcatSmoking}
\end{figure}

A critical advancement in this framework is the treatment of any given CLD as a \textit{signed category}, where:
\begin{itemize}
    \item Objects in the signed category represent variables in the CLD.
    \item Morphisms (arrows) represent causal relationships between variables.
    \item Each morphism is assigned a polarity ($+$ or $-$).
    \item Paths within the diagram are composed according to the typical composition of morphisms of categories, and the polarities for the composed paths are the product of the polarities of their constituent links.
\end{itemize}

In the previously introduced method of representing a CLD in Section \ref{Subsection:closedDiagrams}, a CLD can be thought of as a signed graph (Figure~\ref{figexCLD}). That method effectively represents the link polarities of CLDs. By contrast, the advanced structure presented here uses a signed \textit{category} to represent a CLD. This framework enables the representation of both links (primitive links) polarities and the polarities of composed paths (implied links), thereby allowing the representation of induced polarities for pathways of causal loops, including for feedback loops \cite{sterman2000business}. In this framework, the category representing all CLDs is a \textit{category of signed categories}.

Figure \ref{figexsignedcatSmoking} shows an example of representing the smoking causal loop diagram using the advanced framework of signed categories. In visual representations, primitive links --- direct connection between two variables --- are typically denoted in blue, while we represent in magenta implied links --- indirect causal relationship that arises from a sequence of primitive links. Feedback loops are represented in red.

Representing CLDs using the signed category structure supports the automatic recognition of the presence of implied links and feedback loops. This advanced framework has broad applications. In this paper, we will introduce two key applications: Identification of feedback loops and detecting simple patterns embedded in complex diagrams.

\subsubsection{Application: Identification of Feedback Loops}

\begin{figure}[!h]
\centering
\includegraphics[width=0.15\textwidth]{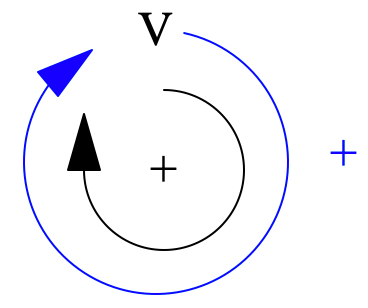}
\caption{The pattern of positive feedback loop.}
\label{figPosLoop}
\end{figure}

\begin{figure}[!h]
\centering
\includegraphics[width=0.15\textwidth]{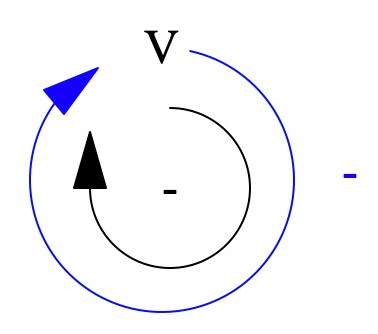}
\caption{The pattern of negative feedback loop.}
\label{figNegLoop}
\end{figure}

Using the advanced framework of signed categories, we can represent a CLD and leverage functor mappings to identify feedback loops. In a manner that recalls --- but greatly generalizes --- how we used a functor from one diagram to another to find patterns in that other diagram, this process involves mapping between the category of pre-defined feedback loop patterns and the CLDs which are represented by signed categories. Figures~\ref{figPosLoop} and~\ref{figNegLoop} illustrate the pre-defined patterns for positive and negative feedback loops, respectively.

A positive feedback loop, as shown in Figure~\ref{figPosLoop}, consists of a dynamic variable $V$ with a positive polarity link to itself. In contrast, a negative feedback loop, depicted in Figure~\ref{figNegLoop}, involves a dynamic variable $V$ with a negative polarity link to itself. 

By applying functor mappings, the pre-defined patterns of positive or negative feedback loops can be identified within a CLD. However, the fact that causal loop diagrams are captured as a category further allows the search for homomorphism to find matches involving composite links within the target category.  As a result, the mapping from the pattern category to the target category results in the detection of feedback loops with net positive or net negative polarities, regardless of the number of links or their specific configuration.  This stands in contrast to pattern finding using the more basic CLDs discussed in Section \ref{PatternIdentification}, which was limited to finding patterns of a pre-defined form.

By leveraging the compositional structure, these patterns can be identified without sensitivity to the naming of variables or the specific sequence of positive and negative polarities, or the number of links involved in the loops -- regardless of size or sequence, a loop will be found as long as the induced polarity for the loop matches the polarity being sought. This flexibility ensures that archetypal structures, such as stabilizing or reinforcing feedback loops, can be consistently recognized in large and complex diagrams.

\subsubsection{Application: Pattern Detection in Elaborate Diagrams}

\begin{figure}[!h]
\centering
\includegraphics[width=0.7\textwidth]{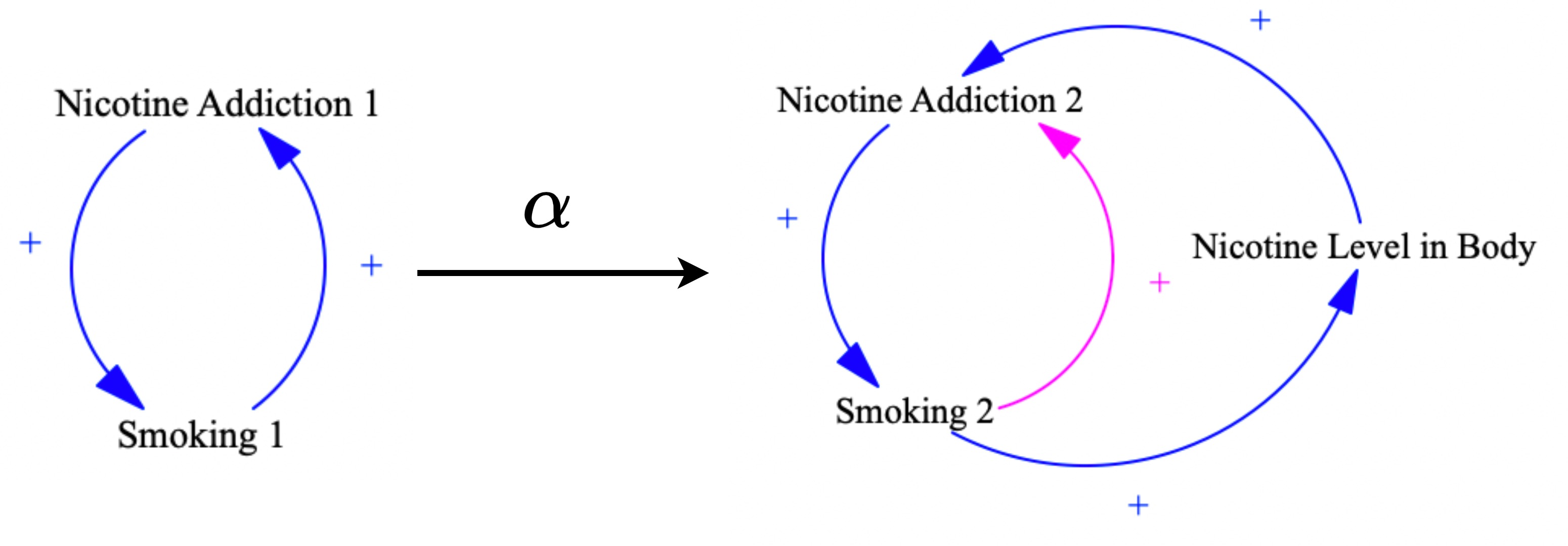}
\caption{An example of detecting a simple pattern (left) embedded within a complex diagram (right).}
\label{figexElaborate}
\end{figure}

Another application, similar in mathematical structure but potentially suited for different situations, is the use of functor mapping to detect simple patterns embedded within complex diagrams.

Figure~\ref{figexElaborate} illustrates an example of detecting a simple pattern in a causal loop diagram (CLD). The simple pattern, represented as ``$Nicotine\,Addiction\,1 \xrightarrow{+} Smoking \xrightarrow{+} Nicotine\,Addiction\,1$", contains two primitive links (blue arrows). This pattern is embedded within a more elaborate CLD, represented as ``$Nicotine\,Addiction\,2 \xrightarrow{+} Smoking\,2 \xrightarrow{+} Nicotine\,Level\,in\,Body \xrightarrow{+} Nicotine\,Addiction\,1$", which contains three primitive polarities (blue arrows). 

The detection of the pattern is achieved via a functor mapping from the simple CLD to the elaborate diagram, denoted as $\alpha$, and where the double arrow indicates the mapping performed by the functor:

\begin{align*}
    Nicotine\,Addiction\,1 \Rightarrow Nicotine\,Addiction\,2, \\
    Smoking\,1 \Rightarrow Smoking\,2 \\
    (Nicotine\,Addiction\,1 \xrightarrow{+} Smoking\,1) \Rightarrow (Nicotine\,Addiction\,2 \xrightarrow{+} Smoking\,2) \\
    ( Smoking\,1\xrightarrow{+} Nicotine\,Addiction\,1) \Rightarrow (Smoking\,2 \xrightarrow{+} Nicotine\,Addiction\,2 )
\end{align*}

Where the arrow ``$Smoking\,2 \xrightarrow{+} Nicotine\,Addiction\,2$" indicates an implied link in the right-hand CLD in the example.

The representation of CLDs as objects within a category of signed categories offers several key advantages:
\begin{itemize}
    \item \textbf{Compositional Analysis:} We can automatically derive higher-order relationships and implied paths within the CLD, enabling deeper insights into system behavior.
    \item \textbf{Pattern Recognition:} The representation can identify recurring structures, such as feedback loops or molecules \cite{molecules, hines1996molecules} in a scalable and robust manner.
    \item \textbf{Flexibility:} The representation of a CLD as a signed category accommodates varying naming conventions and structural configurations while maintaining mathematical rigor.
    \item \textbf{Scalability:} The categorical representation supports large and complex diagrams, enhancing the capacity of the framework to handle real-world systems.
\end{itemize}

By treating causal loop diagrams as signed categories, this approach significantly enhances the flexibility with which we can use them in analysis. It enables rigorous compositional analysis, supports automated identification of feedback loops regardless of size and particular configuration, and systematic exploration of system behavior. These advancements provide a powerful tool for researchers and practitioners seeking to analyze and understand System Dynamics diagrams --- and the dynamical systems that they represent --- in greater depth.

\section{Conclusion and Discussion}

This work opens numerous pathways for advancing System Dynamics modeling by leveraging categorical approaches. Several promising directions and opportunities for future work emerge from this framework, which we outline below.
\begin{itemize}
    \item \textbf{Enhancing Model Representation and Semantics}: By integrating dimensional and unit annotations within models, we can extend the categorical framework to capture essential model characteristics with mathematical rigor. Additionally, this framework allows for connecting the semantics of stock-and-flow diagrams with those from different modeling traditions, such as Petri-net-based models or ordinary differential equations (ODEs) defined separately. This integration enables richer and more versatile modeling capabilities.

    A noteworthy extension involves the use of \textit{temporal sheaves} \cite{temporalSheaves}, which facilitate reasoning about system behavior across multiple temporal scales. Temporal sheaves enable seamless transitions between global and local temporal perspectives, potentially allowing for comprehensive analyses of cumulative flows and dynamic behaviors over extended timeframes.

    \item \textbf{Supporting Model Logic and Provenance}
    
    With an eye towards achieving a better separation of concerns, we aim to better separate model logic from observer processes or quantities associated with reporting derived data. This distinction enhances the clarity of model design and the interpretation of epiphenomenal quantities. Furthermore, advanced categorical structures, such as \textit{toposes}, can support flexible query languages for models. 

    \item \textbf{Expanding to Hybrid and Agent-Based Models}

    While the primary focus of this work is on System Dynamics, a parallel framework for categorical agent-based and hybrid modeling is under development. This framework incorporates declarative representations of agent-based models and associated epiphenomenal observation processes, integrating state charts and stock-and-flow constructs to characterize both discrete and continuous dynamics. This approach replaces traditional reliance on coding of agent behavior in imperative programming languages and message-based inter-agent communication mechanisms with more robust and rigorous mathematical formulations, supporting a more natural combination of modeling paradigms, optimization and analysis.
\end{itemize}

The application of category theory to System Dynamics modeling reveals and formalizes the hidden structure and mathematical properties of System Dynamics diagrams. These diagrams, long recognized as powerful tools for understanding complex systems, benefit significantly from categorical methods in multiple ways, such as by:
\begin{itemize}
    \item Illuminating the mathematical structure underlying multiple types of System Dynamics models.
    \item Facilitating reasoning about models diagrammatically, connecting diagrams of the same type (e.g., causal loop diagrams or stock \& flow diagrams) or across types.
    \item Supporting modularity by enabling the composition of smaller components (e.g., molecules or submodels) into larger, integrated systems.
    \item Improving the transparency, stratification, exploration, comparison and analysis of model structures.
    \item Provide new lenses and languages for understanding model evolution and behavior.
\end{itemize}

The methods and tools developed through this framework are designed to empower System Dynamics teams. Importantly, important subsets of the software tools enabled by these techniques are accessible to users without prior knowledge of category theory, enabling practical benefits such as:
\begin{itemize}
    \item Seamless manipulation and analysis of models.
    \item Mapping between and relating different types of System Dynamics diagrams.
    \item Enhanced collaboration and modularity in team-based modeling efforts.
\end{itemize}

Looking ahead, the ongoing evolution of these toolsets and mathematical foundations holds the potential to transform and help enable System Dynamics modeling. By enabling deeper analysis of model structure, behavior, and the modeling process itself, these methods provide powerful support for addressing pressing global challenges using the full potential of System Dynamics diagrams and teamwork.

\medskip

\section*{Acknowledgments}
The authors wish to offer our gratitude towards others who have helped facilitate the develpment of this work, including Sophie Libkind, Eric Redekopp, Thomas Purdy, Alex Alegre, Nona Sepahrom, Nastaran Jamali, and Medhi Tafarogi. We also thank Jason Brown for helpful comments on an earlier version of this work.

\bibliographystyle{abbrvnat} 
\bibliography{reference}

\end{document}